\documentstyle[12pt,epsf]{article}
\newlength{\pubnumber} 

\catcode`\@=11
\@addtoreset{equation}{section}

\def\section{\@startsection{section}{1}{\z@}{3.5ex plus 1ex minus .2ex}
 {2.3ex plus .2ex}{\large\bf}}
\def\subsection{\@startsection{subsection}{2}{\z@}{2.3ex plus .2ex}
 {2.3ex plus .2ex}{\bf}}

\input epsf
\newfont{\mbm}{msbm10 scaled \magstep 1}
\newfont{\mbmb}{msbm10 scaled \magstep 3}
\newfont{\mbmbb}{msbm10 scaled \magstep 2}
\newfont{\smbm}{msbm10}
\newfont{\ssmbm}{msbm10 at 8pt}
\newfont{\lsmbm}{msbm10 at 14pt}
\def\bbb#1{\hbox{\mbmb #1}}
\def\bbbb#1{\hbox{\mbmbb #1}}
\def\bb#1{\hbox{\mbm #1}}

\def\ssbb#1{\hbox{\ssmbm #1}}

\begin{document}

\begin{titlepage}
\samepage{
\setcounter{page}{1}
\rightline{OUTP--03--04P}
\rightline{\tt hep-th/0302006}
\rightline{May 2002}
\vfill
\begin{center}
 {\Large \bf Open Descendants of \\NAHE--based free fermionic and Type I
$\bbb{Z}_2^n$ models\\} \vfill \vfill {\large
    David J. Clements\footnote{david.clements@new.ox.ac.uk}
        and
        Alon E. Faraggi\footnote{faraggi@thphys.ox.ac.uk}}\\
\vspace{.12in} {\it Theoretical Physics Department, University of
Oxford, Oxford OX1 3NP\\} \vspace{.025in}
\end{center}
\vfill
\begin{abstract}

The NAHE--set, that underlies the realistic free fermionic models,
corresponds to $\bb{Z}_2\times \bb{Z}_2$ orbifold at an enhanced
symmetry point, with $(h_{11},h_{21})=(27,3)$. Alternatively, a
manifold with the same data is obtained by starting with a
$\bb{Z}_2\times \bb{Z}_2$ orbifold at a generic point on the
lattice and adding a freely acting $\bb{Z}_2$ involution. In this
paper we study type I orientifolds on the manifolds that underly
the NAHE--based models by incorporating such freely acting shifts.
We present new models in the Type I vacuum which are modulated by
$\bb{Z}_2^n$ for $n=2,3$.  In the case of $n=2$, the
$\bb{Z}_2\times \bb{Z}_2$ structure is a composite orbifold Kaluza
Klein shift arrangement. The partition function provides a simpler
spectrum with chiral matter.  For $n=3$, the case discussed is a
$\bb{Z}_2$ modulation of the $T^6/\bb{Z}_2 \times \bb{Z}_2$
spectrum.  The additional projection shows an enhanced closed and
open sector with chiral matter.  The brane stacks are
correspondingly altered from those which are present in the
$\bb{Z}_2\times \bb{Z}_2$ orbifold. In addition, we discuss the
models arising from the open sector with and without discrete
torsion.

\end{abstract}
\smallskip}
\end{titlepage}

\setcounter{footnote}{0}

\def\beq{\begin{equation}}
\def\eeq{\end{equation}}
\def\beqn{\begin{eqnarray}}
\def\eeqn{\end{eqnarray}}
\newcommand{\nn}{\nonumber}

\def\no{\noindent }
\def\nolabel{\nonumber }
\def\ie{{\it i.e.}}
\def\eg{{\it e.g.}}
\def\half{{\textstyle{1\over 2}}}
\def\third{{\textstyle {1\over3}}}
\def\quarter{{\textstyle {1\over4}}}
\def\sixth{{\textstyle {1\over6}}}
\def\m{{\tt -}}
\def\p{{\tt +}}

\def\Tr{{\rm Tr}\, }
\def\tr{{\rm tr}\, }

\def\slash#1{#1\hskip-6pt/\hskip6pt}
\def\slk{\slash{k}}
\def\GeV{\,{\rm GeV}}
\def\TeV{\,{\rm TeV}}
\def\y{\,{\rm y}}
\def\SM{Standard--Model }
\def\SUSY{supersymmetry }
\def\SSSM{supersymmetric standard model}
\def\vev#1{\left\langle #1\right\rangle}
\def\l{\langle}
\def\r{\rangle}
\def\o#1{\frac{1}{#1}}

\def\Htw{{\tilde H}}
\def\chibar{{\overline{\chi}}}
\def\qbar{{\overline{q}}}
\def\ibar{{\overline{\imath}}}
\def\jbar{{\overline{\jmath}}}
\def\Hbar{{\overline{H}}}
\def\Qbar{{\overline{Q}}}
\def\abar{{\overline{a}}}
\def\alphabar{{\overline{\alpha}}}
\def\betabar{{\overline{\beta}}}
\def\tautwo{{ \tau_2 }}
\def\thetatwo{{ \vartheta_2 }}
\def\thetathree{{ \vartheta_3 }}
\def\thetafour{{ \vartheta_4 }}
\def\ttwo{{\vartheta_2}}
\def\tthree{{\vartheta_3}}
\def\tfour{{\vartheta_4}}
\def\ti{{\vartheta_i}}
\def\tj{{\vartheta_j}}
\def\tk{{\vartheta_k}}
\def\calF{{\cal F}}
\def\smallmatrix#1#2#3#4{{ {{#1}~{#2}\choose{#3}~{#4}} }}
\def\ab{{\alpha\beta}}
\def\Minv{{ (M^{-1}_\ab)_{ij} }}
\def\bone{{\bf 1}}
\def\ii{{(i)}}
\def\V{{\bf V}}
\def\N{{\bf N}}

\def\b{{\bf b}}
\def\S{{\bf S}}
\def\X{{\bf X}}
\def\I{{\bf I}}
\def\mb{{\mathbf b}}
\def\mS{{\mathbf S}}
\def\mX{{\mathbf X}}
\def\mI{{\mathbf I}}
\def\balpha{{\mathbf \alpha}}
\def\bbeta{{\mathbf \beta}}
\def\bgamma{{\mathbf \gamma}}
\def\bxi{{\mathbf \xi}}

\def\t#1#2{{ \Theta\left\lbrack \matrix{ {#1}\cr {#2}\cr }\right\rbrack }}
\def\C#1#2{{ C\left\lbrack \matrix{ {#1}\cr {#2}\cr }\right\rbrack }}
\def\tp#1#2{{ \Theta'\left\lbrack \matrix{ {#1}\cr {#2}\cr }\right\rbrack }}
\def\tpp#1#2{{ \Theta''\left\lbrack \matrix{ {#1}\cr {#2}\cr }\right\rbrack }}
\def\l{\langle}
\def\r{\rangle}


\def\inbar{\,\vrule height1.5ex width.4pt depth0pt}

\def\IC{\relax\hbox{$\inbar\kern-.3em{\rm C}$}}
\def\IQ{\relax\hbox{$\inbar\kern-.3em{\rm Q}$}}
\def\IR{\relax{\rm I\kern-.18em R}}
 \font\cmss=cmss10 \font\cmsss=cmss10 at 7pt
\def\IZ{\relax\ifmmode\mathchoice
 {\hbox{\cmss Z\kern-.4em Z}}{\hbox{\cmss Z\kern-.4em Z}}
 {\lower.9pt\hbox{\cmsss Z\kern-.4em Z}}
 {\lower1.2pt\hbox{\cmsss Z\kern-.4em Z}}\else{\cmss Z\kern-.4em Z}\fi}

\def\AEF{A.E. Faraggi}
\def\NPB#1#2#3{{\it Nucl.\ Phys.}\/ {\bf B#1} (#2) #3}
\def\PLB#1#2#3{{\it Phys.\ Lett.}\/ {\bf B#1} (#2) #3}
\def\PRD#1#2#3{{\it Phys.\ Rev.}\/ {\bf D#1} (#2) #3}
\def\PRL#1#2#3{{\it Phys.\ Rev.\ Lett.}\/ {\bf #1} (#2) #3}
\def\PRT#1#2#3{{\it Phys.\ Rep.}\/ {\bf#1} (#2) #3}
\def\MODA#1#2#3{{\it Mod.\ Phys.\ Lett.}\/ {\bf A#1} (#2) #3}
\def\IJMP#1#2#3{{\it Int.\ J.\ Mod.\ Phys.}\/ {\bf A#1} (#2) #3}
\def\nuvc#1#2#3{{\it Nuovo Cimento}\/ {\bf #1A} (#2) #3}
\def\RPP#1#2#3{{\it Rept.\ Prog.\ Phys.}\/ {\bf #1} (#2) #3}
\def\JHEP#1#2#3{ {\it JHEP } {\bf #1} (#2)  #3}
\def\etal{{\it et al\/}}

\hyphenation{su-per-sym-met-ric non-su-per-sym-met-ric}
\hyphenation{space-time-super-sym-met-ric}
\hyphenation{mod-u-lar mod-u-lar--in-var-i-ant}


\setcounter{footnote}{0}

\section{Introduction}

Important progress has been achieved in recent years in the basic
understanding of string theory. It is now believed that the
different string theories in ten dimensions, together with eleven
dimensional supergravity, are limits of a single more fundamental
theory, traditionally called M--theory. The question remains,
however, how to relate these advances to experimental data. In
this context some efforts have been directed at the construction
of phenomenologically viable type I string vacua \cite{typeiphun},
and nonperturbative M--theory vacua based on compactifications of
11 dimensional supergravity on CY$\times S_1/\bb{Z}_2$
\cite{horavawit,donagi,fgi} or on manifolds with $G_2$ holonomy
\cite{g2mani}.

These perturbative string constructions, however, do not yet
utilize the new M--theory picture of string theories.
The question remains how to employ this new
understanding for phenomenological studies.
In the context of M--theory
the true fundamental theory of nature should have some
nonperturbative realization. However, at present all we know
about this more basic theory are its perturbative string limits.
Therefore, we should regard these theories as
providing tools to probe the properties of the fundamental nonperturbative
vacuum in the different limits.
Each of the perturbative string
limits may therefore exhibit some properties of the
true vacuum, but it may well be that none
can characterize the vacuum completely.
In this view it is likely that all limits
will need to be used to isolate the true M--theory
vacuum. In this respect it may well
be that different perturbative string limits may provide more useful
means to study different properties of the true nonperturbative vacuum.
This suggests the following
approach to exploration of M--theory phenomenology. Namely, the
true M--theory vacuum has some nonperturbative realization that at
present we do not know how to formulate. This vacuum is at finite
coupling and is located somewhere in the space of M--theory
vacua. The properties of the true vacuum can
however be probed in the perturbative string limits. We may
hypothesize that in any of these limits one still needs to compactify
to four dimensions. Namely, that the true M--theory vacuum
can still be formulated with four non--compact and all the other
dimensions are compact. Suppose then that in some of the limits
we are able to identify a specific class of compactifications
that possess appealing phenomenological properties. The new
M--theory picture suggests that we can then explore the possible
properties of the M--theory vacuum by studying compactifications
of the other perturbative string limits on the same class of
compactifications.

In particular, we can probe those properties that pertain to the
observed experimental and cosmological data, and by using the
low energy effective field theory parameterization.
One of these properties, indicated
by the observed data, is the embedding
of the Standard Model matter states in the chiral ${\bf 16}$
representation
of $SO(10)$.
Thus, we may demand the existence of
a viable perturbative string limit which preserve this embedding.
The only perturbative string limit which enables the $SO(10)$
embedding of the Standard Model spectrum is the heterotic $E_8\times E_8$
string. The reason being that only this limit produces the
spinorial 16 representation in the perturbative massless spectrum.
Therefore, if we would like to preserve the $SO(10)$ embedding of the
Standard Model spectrum, the M--theory limit which we should use is
the perturbative heterotic string \cite{heterotic}.
In this respect it may well
be that other perturbative string limits may provide more useful
means to study different properties of the true nonperturbative vacuum,
such as dilaton and moduli stabilization \cite{casasabel}.

Pursuing this point of view, a class of realistic string models
that preserve the $SO(10)$ embedding of the Standard Model
spectrum are the NAHE--based free fermionic models. This
formulation enables detailed studies at fixed points in the moduli
space, and the models under consideration correspond to
$\bb{Z}_2\times \bb{Z}_2$ orbifold compactifications with
additional Wilson lines\footnote{It is in general anticipated that
the different formulations of string compactifications to four
dimensions do not represent different physics and are related,
even if the dictionary is not always known.}. Many of the
encouraging phenomenological characteristics of the realistic free
fermionic models are rooted in the underlying $\bb{Z}_2\times
\bb{Z}_2$ orbifold structure, including the three generations
arising from the three twisted sectors, and the canonical SO(10)
embedding for the weak hyper-charge. We may therefore regard the
phenomenological success of the free fermionic models as
highlighting a specific class of compactified manifolds.

Given the specific class of compactified manifolds highlighted by
NAHE--based free fermionic models, the line of approach
to phenomenological studies of M--theory
that we pursue here is to compactify
other perturbative string limits on the same manifolds. It is then
hoped that these studies will elucidate other properties of these
realistic models. This is the line of thought that was pursued in
ref. \cite{fgi} where compactification of Horava--Witten theory to
four dimensions on manifolds that are related to the free
fermionic models were studied.

Pursuing this approach we study in this paper orientifolds of type
IIB string theory on the manifolds that are related to the free
fermionic models. The geometric manifold that underlies the free
fermionic models is a $\bb{Z}_2\times \bb{Z}_2$ orbifold at an
enhanced symmetry point in the Narain moduli space. At the free
fermionic point the Narain lattice arising from the six
compactified dimensions is enhanced from $U(1)^6$ to $SO(12)$. The
$\bb{Z}_2\times \bb{Z}_2$ orbifold projection of this lattice then
yields a manifold with $(h_{11},h_{21})=(27,3)$. On the other hand
a $\bb{Z}_2\times \bb{Z}_2$ orbifold projection at a generic point
in the moduli space yields a manifold with
$(h_{21},h_{11})=(51,3)$. We refer to the later as $X_1$ and to
the former as $X_2$. These two manifolds can alternatively be
connected by adding a freely acting shift to $X_1$, which reduces
the number of twisted fixed points by 1/2. Orientifolds of
$\bb{Z}_2\times \bb{Z}_2$ orbifolds were studied in ref.
\cite{z2z2orient}. To advance these studies toward nonperturbative
studies of the free fermionic models we therefore extend this
analysis by including the freely acting shift that connects the
$X_1$ and $X_2$ manifolds.

\setcounter{footnote}{0}
\section{Realistic free fermionic models - general structure}\label{gs}
In this section we recapitulate the main structure of
the realistic free fermionic models.
The notation
and details of the construction of these
models are given elsewhere \cite{rffm}.
In the free fermionic formulation \cite{fff} of the heterotic string
in four dimensions a model is specified in terms of boundary
condition basis vectors and one--loop GSO phases.
The physical spectrum is obtained by applying the generalized GSO projections.
The boundary condition basis defining a typical
realistic free fermionic heterotic string models is
constructed in two stages.
The first stage consists of the NAHE set,
which is a set of five boundary condition basis vectors,
$\{{\bf1},S,b_1,b_2,b_3\}$ \cite{nahe}.
The gauge group after imposing the GSO projections induced
by the NAHE set is $SO(10)\times SO(6)^3\times E_8$
with $N=1$ supersymmetry.
At the level of the NAHE set the sectors $b_1$, $b_2$ and $b_3$
produce 48 multiplets, 16 from each, in the $16$
representation of $SO(10)$. The states from the sectors $b_j$
are singlets of the hidden $E_8$ gauge group and transform
under the horizontal $SO(6)_j$ $(j=1,2,3)$ symmetries.
This structure is common to all the realistic free fermionic models.

The second stage of the
basis construction consists of adding to the
NAHE set three (or four) additional boundary condition basis vectors,
typically denoted by $\{\alpha,\beta,\gamma\}$.
These additional basis vectors reduce the number of generations
to three chiral generations, one from each of the sectors $b_1$,
$b_2$ and $b_3$, and simultaneously break the four dimensional
gauge group. The assignment of boundary conditions
breaks $SO(10)$ to one of its subgroups \cite{rffm}.
Similarly, the hidden $E_8$ symmetry is broken to one of its
subgroups by the basis vectors which extend the NAHE set.
The flavor $SO(6)^3$ symmetries in the NAHE--based models
are broken to flavor $U(1)$ symmetries.
The three additional basis vectors $\{\alpha, \beta, \gamma\}$ differ
between the models and there exists a large number of viable three
generation models in this class.

 From the preceding discussion it follows that the underlying
$\bb{Z}_2\times \bb{Z}_2$ orbifold structure is common to all the
three generation free fermionic models. This is the structure that
we will exploit in trying to elevate the study of these models
across the strong--weak duality barrier. In this respect our aim
is to explore which of the structures of these models is preserved
in the nonperturbative domain. We should note that a priori -- we
have no clue -- and therefore the analysis is purely exploratory.

The correspondence of the NAHE--based free fermionic models
with the orbifold construction is illustrated
by extending the NAHE set, $\{{\bf1},S,b_1,b_2,b_3\}$, by one additional
boundary condition basis vector \cite{foc}, $\xi_1$.
With a suitable choice of the GSO projection coefficients the
model possess an $SO(4)^3\times E_6\times U(1)^2\times E_8$ gauge group
and $N=1$ space--time supersymmetry. The matter fields
include 24 generations in the 27 representations of
$E_6$, eight from each of the sectors $b_1\oplus b_1+\xi_1$,
$b_2\oplus b_2+\xi_1$ and $b_3\oplus b_3+\xi_1$.
Three additional 27 and $\overline{27}$ pairs are obtained
from the Neveu--Schwarz $\oplus~\xi_1$ sector.

To construct the model in the orbifold formulation one starts
with a model compactified on a flat torus with nontrivial background
fields \cite{Narain}.
The subset of basis vectors
\beq
\{{\bf1},S,\xi_1,\xi_2\},
\label{neq4set}
\eeq
with $\xi_2={\bf1}+b_1+b_2+b_3$,
generates a toroidally-compactified model with $N=4$ space--time
supersymmetry and $SO(12)\times E_8\times E_8$ gauge group.
The same model is obtained in the geometric (bosonic) language
by constructing the background fields which produce
the $SO(12)$ lattice. The metric of the six-dimensional compactified
manifold is taken as the Cartan matrix of $SO(12)$,
and the antisymmetric tensor is given by
\beq
B_{ij}=\cases{
G_{ij}&;\ $i>j$,\cr
0&;\ $i=j$,\cr
-G_{ij}&;\ $i<j$.\cr}
\label{bso12}
\eeq
When all the radii of the six-dimensional compactified
manifold are fixed at $R_I=\sqrt2$, it is seen that the
left-- and right--moving momenta
$
P^I_{R,L}=[m_i-{1\over2}(B_{ij}{\pm}G_{ij})n_j]{e_i^I}^*
$
reproduce all the massless root vectors in the lattice of
$SO(12)$. Here $e^i=\{e_i^I\}$ are six linearly-independent
vectors normalized: $(e_i)^2=2$.
The ${e_i^I}^*$ are dual to the $e_i$, with
$e_i^*\cdot e_j=\delta_{ij}$.

Adding the two basis vectors $b_1$ and $b_2$ to the set
(\ref{neq4set}) corresponds to the $\bb{Z}_2\times \bb{Z}_2$
orbifold model with standard embedding. Starting from the Narain
model with $SO(12)\times E_8\times E_8$ symmetry~\cite{Narain},
and applying the $\bb{Z}_2\times \bb{Z}_2$ twisting on the
internal coordinates, reproduces the spectrum of the free-fermion
model with the six-dimensional basis set
$\{{\bf1},S,\xi_1,\xi_2,b_1,b_2\}$. The Euler characteristic of
this model is 48 with $h_{11}=27$ and $h_{21}=3$.

It is noted that the effect of the additional basis vector $\xi_1$
is to separate the gauge degrees of freedom {}from the internal
compactified degrees of freedom. In the realistic free fermionic
models this is achieved by the vector $2\gamma$ \cite{foc}, which
breaks the $E_8\times E_8$ symmetry to $SO(16)\times SO(16)$. The
$\bb{Z}_2\times \bb{Z}_2$ twisting breaks the gauge symmetry to
$SO(4)^3\times SO(10)\times U(1)^3\times SO(16)$. The orbifold
twisting still yields a model with 24 generations, eight from each
twisted sector, but now the generations are in the chiral 16
representation of $SO(10)$, rather than in the 27 of $E_6$. The
same model can be realized with the set
$\{{\bf1},S,\xi_1,\xi_2,b_1,b_2\}$, by projecting out the
$16\oplus{\overline{16}}$ from the sector $\xi_1$ by taking \beq
c{\xi_1\choose \xi_2}\rightarrow -c{\xi_1\choose \xi_2}.
\label{changec} \eeq This choice also projects out the massless
vector bosons in the 128 of $SO(16)$ in the hidden-sector $E_8$
gauge group, thereby breaking the $E_6\times E_8$ symmetry to
$SO(10)\times U(1)\times SO(16)$. The freedom in eq.
({\ref{changec}) correspond to a discrete torsion in the toroidal
orbifold model. At the level of the $N=4$ Narain model generated
by the set (\ref{neq4set}), we can define two models, $Z_+$ and
$Z_-$, depending on the sign of the discrete torsion in eq.
(\ref{changec}). One model, say $Z_+$, produces the $E_8\times
E_8$ model, whereas the second, say $Z_-$, produces the
$SO(16)\times SO(16)$ model. However, the $\bb{Z}_2\times
\bb{Z}_2$ twists act identically in the two models, and their
physical characteristics differ only due to the discrete torsion
eq. (\ref{changec}).

This analysis confirms that the $\bb{Z}_2\times \bb{Z}_2$ orbifold
on the $SO(12)$ Narain lattice is indeed at the core of the
realistic free fermionic models. However, this orbifold model
differs from the $\bb{Z}_2\times \bb{Z}_2$ orbifold on
$T_2^1\times T_2^2\times T_2^3$ with $(h_{11},h_{21})=(51,3)$. In
ref. \cite{befnq} it was shown that the two models are connected
by adding a freely acting twist or shift to the $X_1$ model. Let
us first start with the compactified $T^1_2\times T^2_2\times
T^3_2$ torus parameterized by three complex coordinates $z_1$,
$z_2$ and $z_3$, with the identification \beq z_i=z_i +
1~~~~~~~~~~;~~~~~~~~~~z_i=z_i+\tau_i \label{t2cube} \eeq where
$\tau$ is the complex parameter of each $T_2$ torus. With the
identification $z_i\rightarrow-z_i$, a single torus has four fixed
points at \beq z_i=\{0,1/2,\tau/2,(1+\tau)/2\}. \label{fixedtau}
\eeq With the two $\bb{Z}_2$ twists \beqn &&
\alpha:(z_1,z_2,z_3)\rightarrow(-z_1,-z_2,~~z_3)\cr &&
\beta:(z_1,z_2,z_3)\rightarrow(~~z_1,-z_2,-z_3), \label{alphabeta}
\eeqn there are three twisted sectors in this model, $\alpha$,
$\beta$ and $\alpha\beta=\alpha\cdot\beta$, each producing 16
fixed tori, for a total of 48. Adding to the model generated by
the $\bb{Z}_2\times \bb{Z}_2$ twists in (\ref{alphabeta}), the
additional shift \beq
\gamma:(z_1,z_2,z_3)\rightarrow(z_1+{1\over2},z_2+{1\over2},z_3+{1\over2})
\label{gammashift} \eeq produces again a fixed tori from the three
twisted sectors $\alpha$, $\beta$ and $\alpha\beta$. The product
of the $\gamma$ shift in (\ref{gammashift}) with any of the
twisted sectors does not produce any additional fixed tori.
Therefore, this shift acts freely. Under the action of the
$\gamma$ shift, half the fixed tori from each twisted sector are
paired. Therefore, the action of this shift is to reduce the total
number of fixed tori from the twisted sectors by a factor of
$1/2$, with $(h_{11},h_{21})=(27,3)$. This model therefore
reproduces the data of the $\bb{Z}_2\times \bb{Z}_2$ orbifold at
the free-fermion point in the Narain moduli space.

We noted above that the freely acting shift (\ref{gammashift}),
added to the ${\bb{Z}}_2\times {\bb{Z}}_2$ orbifold at a generic
point of $T_2^1\times T_2^2\times T_2^3$, reproduces the data of
the ${\bb{Z}}_2\times {\bb{Z}}_2$ orbifold acting on the SO(12)
lattice. This observation does not prove, however, that the vacuum
which includes the shift is identical to the free fermionic model.
While the massless spectrum of the two models may coincide their
massive excitations, in general, may differ. The matching of the
massive spectra is examined by constructing the partition function
of the ${\bb{Z}}_2\times {\bb{Z}}_2$ orbifold of an SO(12)
lattice, and subsequently that of the model at a generic point
including the shift. In effect since the action of the
${\bb{Z}}_2\times {\bb{Z}}_2$ orbifold in the two cases is
identical the problem reduces to proving the existence of a freely
acting shift that reproduces the partition function of the SO(12)
lattice at the free fermionic point. Then since the action of the
shift and the orbifold projections are commuting it follows that
the two ${\bb{Z}}_2\times {\bb{Z}}_2$ orbifolds are identical.

On the compact coordinates there are actually three inequivalent ways
in which the shifts
can act. In the more familiar case, they simply translate a generic point
by half the
length of the circle. As usual, the presence of windings in string
theory allows shifts on the T-dual circle, or even asymmetric ones, that
act both on the circle and on its dual. More concretely, for a circle of
length $2 \pi R$, one can have the following possibilities \cite{vwaaf}:
\beqn
A_1\;:&& X_{\rm L,R} \to X_{\rm L,R} + {\textstyle{1\over 2}} \pi R \,,
\nonumber \\
A_2\;:&& X_{\rm L,R} \to X_{\rm L,R} + {\textstyle{1\over 2}} \left(
\pi R \pm {\pi \alpha ' \over R} \right) \,,
\nonumber \\
A_3\;:&& X_{\rm L,R} \to X_{\rm L,R} \pm {\textstyle{1\over 2}}
{\pi \alpha' \over R} \,. \label{a1a2a3} \eeqn There is an
important difference between these choices: while $A_1$ and $A_3$
can act consistently on any number of coordinates, level-matching
requires instead that $A_2$ acts on (mod) four real coordinates.
By studying the respective partition function one finds
\cite{partitions} that the shift that reproduces the $SO(12)$
lattice at the free fermionic point in the moduli space is
generated by the ${\bb{Z}}_2\times {\bb{Z}}_2$ shifts \beqn g\;: &
& (A_2 , A_2 ,0 ) \,,
\nonumber \\
h\;: & & (0, A_2 , A_2 ) \,, \label{gfh} \eeqn where each $A_2$
acts on a complex coordinate. It is then shown that the partition
function of the SO(12) lattice is reproduced. at the self-dual
radius, $R_i = \sqrt{\alpha '}$. On the other hand, the shifts
given in Eq. (\ref{gammashift}), and similarly the analogous
freely acting shift given by $(A_3,A_3,A_3)$, do not reproduce the
partition function of the $SO(12)$ lattice. Therefore, the shift
in eq. (\ref{gammashift}) does reproduce the same massless
spectrum and symmetries of the ${\bb{Z}}_2\times {\bb{Z}}_2$ at
the free fermionic point, but the partition functions of the two
models differ! Thus, the free fermionic ${\bb{Z}}_2\times
{\bb{Z}}_2$ is realized for a specific form of the freely acting
shift given in eq. (\ref{gfh}). However, all the models that are
obtained from $X_1$ by a freely acting ${\bb{Z}}_2$-shift have
$(h_{11},h_{21})=(27,3)$ and hence are connected by continuous
extrapolations. The study of these shifts in themselves may
therefore also yield additional information on the vacuum
structure of these models and is worthy of exploration.

Despite its innocuous appearance the connection between $X_1$ and
$X_2$ by a freely acting shift has the profound consequence of
making the $X_2$ manifold non--simply connected, which allows the
breaking of the SO(10) symmetry to one of its subgroups. Thus, we
can regard the utility of the free fermionic machinery as singling
out a specific class of $\bb{Z}_2\times \bb{Z}_2$ compactified
manifolds. In this context the freely acting shift has the crucial
function of connecting between the simply connected covering
manifold to the non-simply connected manifold. Precisely such a
construction has been utilized in \cite{donagi,fgi} to construct
non-perturbative vacua of heterotic M-theory. In the next section
we turn to study open descendants of $\bb{Z}_2\times \bb{Z}_2$
orbifolds that incorporate such freely acting shifts.

\setcounter{footnote}{0}


\section{$\bbbb{Z}_2\times \bbbb{Z}_2$ Model With Composite Shift Orbifold
Generators}


To illustrate the effects of the freely acting shifts of the type
in eq. (\ref{gammashift}) on the open descendants, we start with a
simpler example of a $\bb{Z}_2$ orbifold, $g$ and an additional
freely acting shift $h$. The action of $g$ and $h$ and their
products is given in eq.
(\ref{eqn:CompositeGenerators})\footnote{This model was analyzed
in collaboration with Carlo Angelantonj and Emilian Dudas.}.

The $\bb{Z}_2\times \bb{Z}_2$ generators have both an action on
the string coordinates (as a parity projection), and the topology
of the internal directions, in that they break $T^6$ to
$T^2_{45}\times T^2_{67}\times T^2_{89}$, with subscripts
referring to the 2-tori directions.  As such, the original type
IIB theory is projected using
\begin{eqnarray}\label{eqn:CompositeGenerators}
g &=& (1~,~1;-1~,-1;-1,-1~),\nn\\
h &=& (A_1,1;~A_1~,~1~;~~1~,~1~), \nn\\
f &=& (A_1,1;-A_1,-1;-1,-1),
\end{eqnarray}
for $A_1$ defined in (\ref{a1a2a3}).  The generators,
(\ref{eqn:CompositeGenerators}) illustrates the shift action on
only one of the coordinates of the relevant torus. The orbifolds
act on all coordinates within a given torus to provide four fixed
points.

This is an interesting model that has a $\bb{Z}_2\times \bb{Z}_2$
structure while preserving ${\cal N}= 2$ supersymmetry. The choice
of generators that has at least two with shift operators, has the
effect of shifting elements of a matrix $M$ (which encodes the
positions of the fixed points)
\begin{eqnarray}\label{eqn:MatrixM}
Tr h Mq^{L_0}q^{\bar{L}_0}
\end{eqnarray}
to the off diagonal positions in the torus amplitude. This implies
that the independent orbit diagrams (those not related by modular
transformation) no longer contribute to the torus amplitude. This
takes away the consideration of a sub class of models associated
with a sign freedom.  As will be shown, sign changes arising from
the discrete torsion terms will necessarily change the charge of
the brane that they couple to, in addition to fundamentally
changing the partition function for the overall sign of the
product of signs $\epsilon_k=\pm 1$.

The way the modulating group generators are written with composite
shift operators, has a twofold effect, firstly it will necessarily
force the number of distinct $D5$ embedding types to become only
one (in this case, the first torus will provide the $D5$ physics).
This happens because the lifting of lattice states for a given
direction by the action of a shift operation forces the tadpole
condition to eliminate the corresponding brane, as discussed also
in \cite{AADS}. Secondly, the particular arrangement of the
shifted directions will allow for an interesting geometrical
configuration of the $O$-planes in the Klein amplitude.

>From (\ref{eqn:CompositeGenerators}), it is appreciated that the
vacuum is left with 2 degrees of freedom in the $R$-$R$ sector.
This is easily seen by appreciating that an orbifold projection
acts as a $\pi$ rotation under $SU(2)$ generators on the moduli.
Which in this case, the orbifold operation effects only two of the
internal directions. The model thus has an ${\cal N}= 2$ supersymmetry.

The ${\cal N}=2$ character set is derived from the breaking of the
original SO(8) lightcone characters $O_8$, $V_8$, $C_8$ and $S_8$
to supersymmetric representations involving $O_4$, $V_4$, $C_4$
and $S_4$.  The type I constructions are discussed in detail in
\cite{CAAS}. In this case, the supersymmetric world sheet fermion
contributions are written as
\begin{table}[!h]
\begin{center}
\begin{tabular}{ll}
$Q_o=V_4O_4-C_4C_4$, & $Q_v=O_4V_4-S_4S_4$ \\
$Q_s=O_4C_4-S_4O_4$, & $Q_c=V_4S_4-C_4V_4$.
\end{tabular}
\end{center}
\end{table}

The $SO(2n)$ characters are,
\begin{table}[!h]
\begin{center}
\begin{tabular}{ll}
$O_{2n}=\frac{1}{2\eta^n}\big( \theta^n_3+\theta^n_4
\big)$, &
$V_{2n}=\frac{1}{2\eta^n}\big( \theta^n_3-\theta^n_4 \big)$, \\
$S_{2n}=\frac{1}{2\eta^n}\big( \theta^n_2+i^{-n}\theta^n_1 \big)$
& $C_{2n}=\frac{1}{2\eta^n}\big( \theta^n_2-i^{-n}\theta^n_1
\big)$.
\end{tabular}
\end{center}
\end{table}\label{eqn:exp}\\
with the Dedekind eta function
\begin{eqnarray}
\eta=q^{\frac{1}{24}}\prod_{n=1}^{\infty}\big(1-q^n\big),\nn
\end{eqnarray}
their respective conformal weights are $0$, $\frac{1}{2}$,
$\frac{n}{8}$ and $\frac{n}{8}$. Here, these are representations
of a scalar, a vector, and spinors of opposite chirality. The
theta functions originate from the $NS$ and $R$ sectors and are
defined by
\begin{eqnarray}
\theta \left[ \matrix{\chi \cr \phi \cr } \right] =\sum_n
q^{\frac{1}{2}(n+\chi)^2}e^{2\pi i(n+\chi)\phi}
\end{eqnarray}
where $\chi$ and $\phi$ take the values $\frac{1}{2}$ $(NS)$ and
$0$ $(R)$, the labelled theta functions are then defined as,
\begin{eqnarray}
\theta_1 = \left[ \matrix{ \frac{1}{2} \cr \frac{1}{2} \cr }
\right], \quad \theta_2 = \left[ \matrix{ \frac{1}{2} \cr 0 \cr }
\right], \quad \theta_3 = \left[ \matrix{ 0 \cr 0 \cr }
\right]\quad{\rm and}\quad \theta_4 = \left[ \matrix{ 0 \cr
\frac{1}{2} \cr } \right].
\end{eqnarray}

We will use a compact notation for the lattice modes arising from
the compactification on a torus.  In a given direction of a torus,
one has a lattice of the form
\begin{eqnarray}\label{eqn:compact}
\Lambda_{m+a,n+b}=\frac{q^{\frac{\alpha\prime}{4}{\big{(}\frac{(m+a)}
{R}+\frac{(n+b)R}{\alpha\prime}\big{)}}^2}\bar{q}^{\frac{\alpha\prime}{4}
{\big{(}\frac{(m+a)}{R}-\frac{(n+b)R}{\alpha\prime}\big{)}}^2}}{\eta(q)
\eta(\bar{q})}.\nn
\end{eqnarray}

Where the values $a$ and $b$ are in anticipation of the effect of
winding or momentum shifts under $S$ transformation.

To obtain modular invariance under $SL(2,\bb{Z})$, as required by
the topology of the one loop string amplitude, one must perform
$S$ and $T$ transforms, the generators of this group, which act on
the complex torus covering parameter $\tau$ as
\begin{eqnarray}\label{eqn:tran}
S:\tau\rightarrow-\frac{1}{\tau} && \Rightarrow(a,b)\rightarrow(b,a^{-1})\nn\\
T:\tau\rightarrow\tau+1 && \Rightarrow(a,b)\rightarrow(a,a b)\nn\\
\end{eqnarray}
Here, $a$ and $b$ label the orbifold/twist operations that are
placed on two sides of the torus sheet.  The full orbit
configuration of these operators is described for the
$\bb{Z}_2\times\bb{Z}_2$ case in \ref{app:Blocks}.
\begin{table}[!ht]
\begin{center}
\begin{tabular}{|c|c|c|}\hline
Lattice & ${\cal A}$ & ${\cal K}~$and$~{\cal M}$ \\\hline $P_m$ &
$W_n$ &
$W_{2n}$\\
$(-1)^mP_{m+\frac{1}{2}}$ & $(-1)^nW_{n+\frac{1}{2}}$ &
$(-1)^nW_{2n+1}$\\
$P_{m+\frac{1}{2}}$ & $(-1)^nW_n$ & $(-1)^nW_{2n}$\\
$(-1)^mP_m$ & $W_{n+\frac{1}{2}}$ & $W_{2n+1}$ \\\hline
\end{tabular}
\caption{Lattice $S$ transforms}\label{tab:Stransforms}
\end{center}
\end{table}

Here, ${\cal A}$, ${\cal K}$ and ${\cal M}$ are the annulus Klein
and Mobius contributions, and the relevant terms can be seen by
using the appropriate form for the measure parameter in each
case\footnote{The shift in mass by applying an $S$ transformation
on terms involving a phase is illustrated in appendix
\ref{app:massshift}}.  $P$ and $W$ are the restriction of
$\Lambda_{m,n}$ to pure Kaluza-Klein ($P$) or winding ($W$) modes.
The further notation of $P_o$ and $P_e$ (and similarly for the
winding sums) are the restriction of the counting to even or odd
modes only.  In the case of odd lattices, the convention should
not be taken to be correlated with labels on the fermionic
contributions. The action on these lattice modes for $S$
transformations are as in table \ref{tab:Stransforms}.  The action
of $T$ on the lattices are
\begin{eqnarray}\label{eqn:Ttransforms}
\begin{array}{lcl}
\Lambda_{m,n}\rightarrow\Lambda_{m,n}, & &
\Lambda_{m,n+\frac{1}{2}}\rightarrow(-1)^m\Lambda_{m,n+\frac{1}{2}}, \\
\Lambda_{m+\frac{1}{2},n}\rightarrow(-1)^n\Lambda_{m+\frac{1}{2},n},
& { \rm and } & \Lambda_{m+\frac{1}{2},n+\frac{1}{2}}\rightarrow
i(-1)^{m+n}\Lambda_{m+\frac{1}{2},n+\frac{1}{2}}.\nn
\end{array}
\end{eqnarray}

Thus the modular invariant torus amplitude is
\begin{eqnarray}
{\cal
T}=&\frac{1}{4}&\bigg\{\big[1+(-1)^{m_1+m_2}\big](\Lambda^1\Lambda^2+
\Lambda^1_{m,n+\frac{1}{2}}\Lambda^2_{m,n+\frac{1}{2}})\Lambda^3|Q_o+Q_v|^2\nn\\
&&+\big[1+(-1)^{m_1}\big]\Lambda^1|Q_o-Q_v|^2{\vline\frac{2\eta}{\theta_2}
\vline}^4\nn\\
&&+16(\Lambda^1+\Lambda^1_{m,n+\frac{1}{2}}){\vline\frac{\eta}{\theta_4}
\vline}^4|Q_s+Q_c|^2\nn\\
&&+16(\Lambda^1+(-1)^{m_1}\Lambda^1_{m,n+\frac{1}{2}}){\vline\frac{\eta}
{\theta_3}\vline}^4|Q_s-Q_c|^2\bigg\}.\nn\\
\end{eqnarray}
With the $A_1$ shift operator, the number of fixed points can be
seen to be halved, it acts on the fixed point coordinates as
\begin{eqnarray}
(0,0;0,0)\rightarrow(0+\frac{1}{2},0;0+\frac{1}{2},0).
\end{eqnarray}
Where the labelling $(x_2,y_2;x_3,y_3)$ defines the collective
fixed point coordinate for the space $T^2_{67}\times T^2_{89}$,
for the values
$\{x_i,y_i|x_i\in\{0,\frac{1}{2}\},y_i\in\{0,\frac{1}{2}\}\}$. The
total number of fixed points without the shift operation is
$16=4\times 4$, which are detailed in table
(\ref{tab:fixedpoints}).
\begin{table}[!ht]
\begin{center}
\begin{tabular}{|llll|}\hline
${(0,0;0,0)}_1$ & ${(0,\frac{1}{2};0,0)}_2$ &
${(\frac{1}{2},0;0,0)}_3$ & ${(\frac{1}{2},\frac{1}{2};0,0)}_4$
\\ ${(0,0;0,\frac{1}{2})}_5$ & ${(0,0;\frac{1}{2},0)}_6$ &
${(0,0;\frac{1}{2},\frac{1}{2})}_7$ & ${(0,\frac{1}{2};0,\frac{1}{2})}_8$\\
${(0,\frac{1}{2};\frac{1}{2},0)}_9$ &
${(0,\frac{1}{2};\frac{1}{2},\frac{1}{2})}_{10}$ &
${(\frac{1}{2},0;0,\frac{1}{2})}_{11}$ &
${(\frac{1}{2},0;\frac{1}{2},0)}_{12}$\\
${(\frac{1}{2},0;\frac{1}{2},\frac{1}{2})}_{13}$ &
${(\frac{1}{2},\frac{1}{2};0,\frac{1}{2})}_{14}$ &
${(\frac{1}{2},\frac{1}{2};\frac{1}{2},0)}_{15}$ &
${(\frac{1}{2},\frac{1}{2};\frac{1}{2},\frac{1}{2})}_{16}$ \\ \hline
\end{tabular}
\caption{Unshifted fixed points}\label{tab:fixedpoints}
\end{center}
\end{table}
The origin of the lattice contributions of the torus amplitude
\begin{eqnarray}
{\cal T}_0=|Q_o|^{2}+|Q_v|^2+8|Q_s|^2+\ldots.
\end{eqnarray}
shows, as expected, 8 fixed points, reduced from 16 within a given
orbifold projection, the independent coordinates of which are as
in table \ref{tab:shiftedFixedPoints}.
\begin{table}[!ht]
\begin{center}
\begin{tabular}{|llll|}\hline
${(0,0;0,0)}_1$ & ${(0,\frac{1}{2};0,0)}_2$ &
${(\frac{1}{2},0;0,0)}_3$ &
${(\frac{1}{2},\frac{1}{2};0,0)}_4$ \\
${(0,0;0,\frac{1}{2})}_5$ & ${(0,0;\frac{1}{2},\frac{1}{2})}_7$ &
${(0,\frac{1}{2};0,\frac{1}{2})}_8$ &
${(0,\frac{1}{2};\frac{1}{2},\frac{1}{2})}_{10}$ \\
\hline
\end{tabular}
\caption{Remaining fixed points}\label{tab:shiftedFixedPoints}
\end{center}
\end{table}

Vertex operators of states flowing in $K$ and $\tilde{{\cal A}}$
will acquire from the torus, by virtue of the action of the shift
in $T_{45}^2$ and $T_{67}^2$, a state projector
\begin{eqnarray}
V=\big[1+(-1)^{m_1+m_2}\big]V_{(T^4/\ssbb{Z}_2)\times T_2}.
\end{eqnarray}

The torus amplitude in the type I setting is accompanied by the
Klein bottle amplitude, which is realized from the type IIB
projection
\begin{eqnarray}
Tr_{IIB}\frac{(1+\Omega)}{2}q^{L_0}{\bar{q}}^{\tilde{L}_0}.
\end{eqnarray}
Where $\Omega$ has the usual definition of the world sheet parity
operator. Terms contributing to the Klein correspond to those
terms in the trace with the $\Omega$ insertion.

$\Omega$ makes an effective identification of the left and right
modes.  As such, orbifold elements acting on the world sheet
bosonic or fermionic oscillators are made ineffective by $\Omega$.
This is easily seen by series expansion of such terms, since the
left and right modes contribute $(-1)^{k+\tilde{k}}$,
$k,\tilde{k}\in \bb{Z}$, the identification then neglects the
orbifold presence.

In a similar fashion, this projection also reduces the lattice
modes to become either pure momentum or pure winding, this
situation is inverted with the assistance of an inserted orbifold
action $\alpha$:
\begin{eqnarray}
\Omega|p_L,p_R>&=&|p_R,p_L>\quad\Rightarrow\quad n=0,\nn\\
\Omega\alpha|p_L,p_R>&=&|-p_R,-p_L>\quad\Rightarrow\quad m=0.
\end{eqnarray}
$\Omega$ has no effect on the twisting operations, since these are
realized as a shift in the oscillator modes. Thus, the Klein
amplitude takes the form,
\begin{eqnarray}
{\cal K}=&\frac{1}{8}&\bigg\{\bigg[\big(1+(-1)^{m_1+m_2}\big)P_1P_2P_3\nn\\
&&+\big(1+(-1)^{m_1}\big)P_1W_2W_3\bigg](Q_o+Q_v)\nn\\
&&+32(Q_s+Q_c)P_1{\bigg(\frac{\eta}{\theta_4}\bigg)}^2\bigg\}
\end{eqnarray}
With corresponding transverse amplitude
\begin{eqnarray}
\tilde{{\cal
K}}=&\frac{2^5}{8}&\bigg\{\bigg[(W_1^eW_2^e+W_1^oW_2^o)W_3^ev_1v_2v_3+
W_1P_2^eP_3^e\frac{v_1}{v_2v_3}\bigg](Q_o+Q_v)\nn\\
&&+2v_1W_1^e(Q_o-Q_v){\bigg(\frac{\eta}{\theta_2}\bigg)}^2)\bigg\}
\end{eqnarray}

Although the $O$-planes present are not indicated explicitly
within amplitudes, there presence and dimension are understood
from the toroidal volumes given by the $v_i$ terms.  $O9$ planes
occupy the entire compact space and so correspond to the term
$v_1v_2v_3$. $O5$ only has a presence in the first of the three
2-tori, and so has a volume term $\frac{v_1}{v_2v_3}$.

The geometry of the O-planes here provide an interesting
realization, they arrange themselves in a diagonal manner due to
the $h$ projection, which is a pure shift. For example, by
performing 2 T-dualities along the 2 directions where the $h$
shift acts in $T_1^2$ and $T_2^2$, the dilaton wave function
\begin{eqnarray}
\phi(y_1,y_2)=\sum_{m_1,m_2}\bigg(cos\big(\frac{m_1y_1}{R_1}\big)
cos\big(\frac{m_2y_2}{R_2}\big)+sin\big(\frac{m_1y_1}{R_1}\big)
sin\big(\frac{m_2y_2}{R_2}\big)\bigg)\phi^{(m_1,m_2)}\nn
\end{eqnarray}
gives access to the positions as in (figure \ref{KFP})

\begin{figure}
\centerline{\epsfxsize 1.0 truein \epsfbox {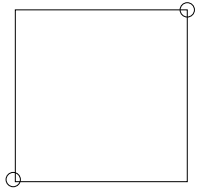}}
\caption{Klein Fixed Point Orientation} \label{KFP}
\end{figure}

By a matter of interpretation, the charges must arrange themselves
as the perfect squares.  This comes from the transverse channel that provides
a tree level coupling between two orientifolds and a closed string.  The cross
terms
then give the mixing of different orientifold types.  The same is true in the
transverse
annulus for brane couplings.

The origin of the lattices in the
transverse Klein amplitude of tori that contribute to the perfect
squares shows their form as
\begin{eqnarray}
\tilde{{\cal
K}}_o&=&\frac{2^5}{8}v_1\bigg\{\big(\sqrt{v_2v_3}+\frac{1}{\sqrt{v_2v_3}}
\big)^2W_1^eQ_o+\big(\sqrt{v_2v_3}-\frac{1}{\sqrt{v_2v_3}}\big)^2W_1^eQ_v\nn\\
&&+\bigg[v_2v_3\big(W_1^oW_2^oW_3^e+W_1^e(W_2^eW_3^e-1)\big)\nn\\
&&+\frac{1}{v_2v_3}\big(W_1^oP_2^eP_3^e+W_1^e(P_2^eP_3^e-1)\big)\bigg](Q_o+Q_v
)\bigg\}
\end{eqnarray}
with massive states shown to illustrate their separate and self
factorization.

The transverse annulus is derived from the states that flow in the
torus, with the restriction to winding (Neumann boundary
conditions) or Kaluza Klein (Dirichlet boundary conditions)
states. The differing boundary conditions, as provided by the
lattice towers, then provides branes of the $D9$ and $D5$ types.

The transverse amplitude is
\begin{eqnarray}\label{eqn:CompositeTransverseAnnulus}
\tilde{{\cal
A}}&=&\frac{2^{-5}}{8}v_1\bigg\{\bigg[N^2v_2v_3(W_1W_2+W_1^{n+\frac{1}{2}}
W_2^{n+\frac{1}{2}})W_3\nn\\
&&+\frac{4D^2}{v_2v_3}W_1P^e_2P_3\bigg](Q_o+Q_v)+4NDW_1(Q_o-Q_v)
{\bigg(\frac{2\eta}{\theta_2}\bigg)}^2\bigg\}\nn\\
&&+\frac{2^{-3}}{8}v_1\bigg\{\bigg[R_N^2(W_1+W_1^{n+\frac{1}{2}})+2R_D^2W_1\bigg]
(Q_s+Q_c){\bigg(\frac{2\eta}{\theta_4}\bigg)}^2\nn\\
&&-4R_NR_DW_1(Q_s-Q_c){\bigg(\frac{\eta}{\theta_3}\bigg)}^2\bigg\}.
\end{eqnarray}
The transverse states (\ref{eqn:CompositeTransverseAnnulus}) then
highlight the $D5$ orientation as in figure \ref{TransA}. The
$D9$'s have been reduced to $D5^\prime$'s by the use of T
dualization on the 4,5,8 and $9^{\rm th}$ coordinates, where there
are two sets of such states located at the origin.  This can be
seen by the integer and half integer massive states that couple to
the $D9$'s in the transverse channel. By performing T dualizations
on these coordinates the $D5$'s are effectively rotated to allow
them to wrap the $T_{89}$ torus. The fixed points relevant to the
$D5$ branes are denoted by circles. The illustration thus shows a
rather standard geometry of the $D5$ branes.  The $D9$ (denoted by
the dashed line) is now a $D5^\prime$ and lies in the $T^2_{67}$
direction.

Some explanation of the numerical coefficients in the above
amplitude is necessary.  In the case of the untwisted terms, one
must satisfy the perfect square structure for the $D5$ and $D9$
terms, as shown more clearly in (\ref{eqn:TransAnnZero}).

The twisted terms are more subtle.  Since such terms effectively
highlight the occupation of branes on the fixed points, their
coefficients must therefore reflect this.  The breaking term $R_N$
corresponds to the effect of the orbifold on the $D9$ brane which
fills all compact and non-compact dimensions. It is wrapped around
all compact dimensions and therefore sees all the fixed points.
The coefficient formula is
\begin{eqnarray}\label{eqn:FixedPointSummary}
\sqrt{\frac{{\rm number~of~fixed~points}}{{\rm
number~of~seen~fixed~points}}}
\end{eqnarray}
$R_N$ thus has the coefficient
\begin{eqnarray}
\sqrt{\frac{16}{16}}\sqrt{v_1}.
\end{eqnarray}

With the volume $v_1$ being provided by the remaining compact
direction that is not acted on by the orbifold element $(+,-,-)$
(and thus has a winding tower in the transverse channel).  The
$D5$ breaking term, $R_D$, involves a brane which wraps only the
first tori, and is transverse to the remaining ones.  Since the
orbifold element $(+,-,-)$ provides fixed points in the second and
third tori, this term therefore has a coefficient of 4, as it does
not see any of these fixed points.

Now, under the identification of the fixed points, one can
categorize the types of brane that see certain fixed points.  All
brane types see the fixed point $(0,0;0,0)$.  So on has the
perfect square
\begin{eqnarray}
\big(R_N\pm 4R_D\big)^2v_1
\end{eqnarray}
where the sign depends on which character they couple to.  For all
other fixed points, $R_D$ does not contribute to the counting
since it only sees $(0,0;0,0)$.  So, the remaining seven fixed
points are taken into account by $R_N$ alone.  There is an overall
factor of 2 that reflects the degeneracy of the original sixteen
fixed points, which is also required for proper particle
interpretation in the direct channel. These details provide the
form for the lattice origin of the transverse annulus as
\begin{eqnarray}\label{eqn:TransAnnZero}
\tilde{{\cal A}}_o&=&\frac{2^{-5}}{8}v_1\bigg\{\bigg(
N\sqrt{v_2v_3}+\frac{2D}{\sqrt{v_2v_3}}
\bigg)^2W_1Q_o+\bigg(N\sqrt{v_2v_3}-\frac{2D}{\sqrt{v_2v_3}}\bigg)^2W_1Q_v\nn\\
&&+\bigg[N^2v_2v_3W_1(W_2W_3-1)+\frac{4D^2}{v_2v_3}W_1(P^e_1P_3-1)\bigg](Q_o+Q_v)
\bigg\}\nn\\
&&+\frac{2^{-5}}{4}v_1\bigg\{\bigg[(R_N-4R_D)^2+7R_N^2\bigg]Q_s+\bigg[(R_N+4R_D)^2
+7R_N^2\bigg]Q_c\bigg\}W_1\nn\\
&&\frac{2^{-5}}{4}v_1\times8R_N^2W_1^{n+\frac{1}{2}}(Q_s+Q_c).
\end{eqnarray}
%
%

\begin{figure}
\centerline{\epsfxsize 2.0 truein \epsfbox {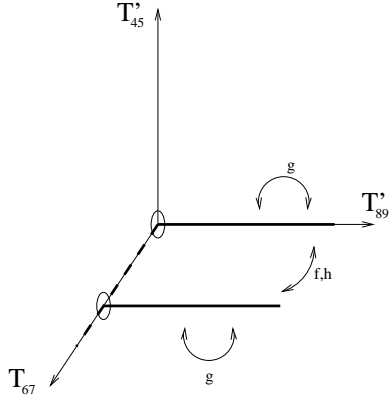}}
\caption{$D5$ and $D5^\prime$ configuration} \label{TransA}
\end{figure}

There is an ambiguity in the form of a sign in the Mobius, it is
chosen so as to allow a consistent tadpole cancellation. This
ambiguity comes from the square root of various coupling terms,
for example, the $D9-O9$ term is given by
\begin{eqnarray}
\tilde{\cal
M}_{D9-O9}=\pm2\times\sqrt{\frac{2^5}{8}}\times\sqrt{\frac{2^{-5}N^2}{8}}.
\end{eqnarray}
Where there is a diagram symmetry factor of $2$ since this a
closed state propagating between a crosscap and $D$-brane
boundary.

The transverse Mobius is then provided as,
\begin{eqnarray}
\tilde{{\cal
M}}&=&-\frac{v_1}{4}\bigg\{\bigg[Nv_2v_3(W_1^eW_2^e+W_1^oW_2^e)W^e_3\nn\\
&&+\frac{2D}{v_2v_3}(W^e_1P^e_2+W^o_1(-1)^{m_2}P^e_2)P_3^e\bigg]
(\hat{Q}_o+\hat{Q}_v)\nn\\
&&+(NW_1+2DW^e_1)(\hat{Q}_o-\hat{Q}_v){\bigg(\frac{2\hat{\eta}}
{\hat{\theta}_2}\bigg)}^2\bigg\}\nn\\
\end{eqnarray}

The corresponding direct channel amplitudes for the annulus and
Mobius are
\begin{eqnarray}
{\cal A}&=&\frac{1}{8}\bigg\{\bigg[N^2\big(1+(-1)^{m_1+m_2}\big)P_1P_2P_3\nn\\
&&+2D^2P_1(W_2+W_2^{n+\frac{1}{2}})W_3\bigg](Q_o+Q_v)\nn\\
&&+4NDP_1(Q_s+Q_c){\bigg(\frac{\eta}{\theta_4}\bigg)}^2+2\bigg[R_N^2P_1^e+
R_D^2P_1\bigg](Q_o-Q_v){\bigg(\frac{2\eta}{\theta_2}\bigg)}^2\nn\\
&&+4R_NR_DP_1(Q_s-Q_c){\bigg(\frac{\eta}{\theta_3}\bigg)}^2\bigg\}\nn\\
\end{eqnarray}
and
\begin{eqnarray}\label{eqn:DirectMobiusSimple}
{\cal M}&=&-\frac{1}{8}\bigg\{\big[N\big(1+(-1)^{m_1+m_2}\big)P_1P_2P_3\nn\\
&&+2D\big(P_1W_2+(-1)^{m_1}P_1W_2^{n+\frac{1}{2}}\big)W_3\big](\hat{Q}_o+
\hat{Q}_v)\nn\\
&&-2(NP^e_1+DP_1)(\hat{Q}_o-\hat{Q}_v){\bigg(\frac{2\hat{\eta}}{\hat{\theta}_2}
\bigg)}^2\bigg\}\nn\\
\end{eqnarray}

In order to extract the tadpole information attention must be paid
to the differing powers of $q$ from the $NS$ and R vacua.  The
Laurant modes
\begin{eqnarray}
L_n=\frac{1}{2}:\sum_n \alpha_{n-l}.\alpha_l:+\frac{1}{2}:\sum_w
(w-\frac{n}{2})\phi_{n-w}\phi_w:+\delta_{n,0}\Delta\nn\\
\end{eqnarray}
for transverse oscillations acquire a total addition of
$-\frac{1}{16}(D-2)$ from the $NS$ sector and 0 from the R. The
zero modes in the amplitude $\tilde{{\cal K}}+\tilde{{\cal
A}}+\tilde{{\cal M}}$, which correspond to ${\mathcal{O}}(q^0)$
terms, give rise to a divergence. Such contributions are tadpole
diagrams and consistency with their cancellation forces a
constraint on the gauge group dimension.  The construction so far
then yields the following tadpole conditions
\begin{eqnarray}
N=32,~2D=32,~R_N=0,~R_D=0.\nn
\end{eqnarray}

The required Chan-Paton parameterization is then
\begin{eqnarray}
N=n+\bar{n},~D=d+\bar{d},~R_N=i(n-\bar{n}),~R_D=i(d-\bar{d}).\nn
\end{eqnarray}
This Chan-Paton charge arrangement shows that the vector multiplet
is not contained in the Mobius.  However, the annulus does, and so
vector multiplet is oriented.  As such the multiplicities have a
unitary interpretation as shown above.

With these representations, one finds that the open sector has the
gauge group breaking (from the character $Q_o$)
\begin{eqnarray}
U(16)_9\times U(16)_5\rightarrow U(16)_9\times U(8)_5\nn,
\end{eqnarray}
under the action of the freely acting shift, and shows the
appropriate gauge couplings as
\begin{eqnarray}
{\cal
A}_o+M_o&=&(n\bar{n}+d\bar{d})Q_o+\frac{1}{2}\big(n(n-1)+\bar{n}
(\bar{n}-1)+d(\bar{d}-1)+\bar{d}(\bar{d}-1)\big)Q_v\nn\\
&&+(n\bar{d}+\bar{n}d)Q_s+(nd+\bar{n}\bar{d})Q_c.
\end{eqnarray}
It is now a simple matter to extract the interesting spectral
content. It has chiral matter in the form of hypermultiplets in
the representations $(120\oplus\overline{120},1)$ and
$(1,28\oplus\overline{28})$ from $Q_v$ as
\begin{eqnarray}
Q_v^h\sim O_2O_2(O_2V_2+V_2O_2)-(S_2S_2+C_2C_2)(S_2S_2+C_2C_2)\nn
\end{eqnarray}
and $(16,\overline{8})$ from $Q_s$ as
\begin{eqnarray}
Q_s^h\sim O_2O_2(C_2S_2+S_2C_2)-(S_2S_2+C_2C_2)O_2O_2.\nn
\end{eqnarray}
%
%


\section{$T^{6}/\bbbb{Z}_{2}^3$ model}


This model is in essence, a generalization of the previous one.
That is that the untwisted states in the parent torus are similar
with the addition of other sectors from an enhanced orbifold
group.  Also, the influence of the momentum shift now extends to
have an effect in all three tori.

The projection that realizes the orbifold structure is represented
as
\begin{eqnarray}
\frac{1}{8}(1+g)(1+f)(1+\delta)\nn
\end{eqnarray}
The $\bb{Z}_2 \times \bb{Z}_2 \times \bb{Z}_2$ generators are
\begin{eqnarray}
&&g=(1,1;-1,-1;-1,-1),\nn\\&&f=(-1,-1;1,1;-1,-1),\nn\\&&\delta=
(A_1,1;A_1,1;A_1,1).\nn
\end{eqnarray}
Unlike the previous case, there are pure orbifold elements present
in the $\bb{Z}_2 \times \bb{Z}_2 \times \bb{Z}_2$ projection,
which will leave trace components on the diagonal of the matrix
$M$ in (\ref{eqn:MatrixM}).  So there will be terms in the torus
amplitude that are not connected by $S$ and $T$ transforms on the
principle orbits $(o,o), (o,g), (o,f)$ and $(o,h)$\footnote{This
is better illustrated in appendix C}. These terms will be realized
in terms of modular orbits that are twisted sectors with different
orbifold insertions, such as $(f,g)$. As such, an ambiguity will
be present in the form of a sign freedom. This will give rise to
models with $(-)$ or without $(+)$ discrete torsion, and
necessitate the study of different classes of models within a
choice of sign (as shown in \cite{CAAS} for the $\bb{Z}_2 \times
\bb{Z}_2$ case without shifts).  The introduction of negative
signs, will also create inconsistencies with the tadpole
conditions that arise from the $NS$ and $R$ sectors.

The torus amplitude results from projecting the type IIB trace as
\begin{eqnarray}
{\cal
T}=\frac{1}{8}TrP_{GSO}(1+g)(1+f)(1+\delta)q^{L_0}\bar{q}^{\tilde{L}_0}.\nn
\end{eqnarray}
with the explicit IIB projection factor of $\frac{1}{2}$ not shown
for brevity.

The models exhibit ${\cal N}=1$ SUSY which results from the
orbifold action on the Ramond sector so as to yield only one
independent fermionic ground state. The torus amplitude results
from the projected trace as
\begin{eqnarray}
{\cal
T}&=&\frac{1}{8}\bigg{\{}|T_{oo}|^2\big{[}{\Lambda^1}_{m,n}{\Lambda^2}_{m,n}
{\Lambda^3}_{m,n}\nn\\
&&+{\Lambda^1}_{m,n+\frac{1}{2}}{\Lambda^2}_{m,n+\frac{1}{2}}
{\Lambda^3}_{m,n+\frac{1}{2}}\big{]}\big{(}1+(-1)^{m_1+m_2+m_3}\big{)}\nn\\
&&+|T_{ok}|^2{\Lambda^k}_{m,n}\big(1+(-1)^{m_k}\big){\vline\frac{2\eta}
{\theta_2}\vline}^4\nn\\
&&+16|T_{ko}|^2\big{(}{\Lambda^k}_{m,n}+{\Lambda^k}_{m,n+\frac{1}{2}}\big{)}
{\vline\frac{\eta}{\theta_4}\vline}^4 \nn\\
&&+16|T_{kk}|^2\big{(}{\Lambda^k}_{m,n}+(-1)^{m_k}
{\Lambda^k}_{m,n+\frac{1}{2}}\big{)}
{\vline\frac{\eta}{\theta_3}\vline}^4 \nn\\
&&+\epsilon(|T_{gh}|^2+|T_{gf}|^2+|T_{fg}|^2+|T_{fh}|^2+|T_{hg}|^2+|T_{hf}|^2)
{\vline\frac{8{\eta}^3}{\theta_2\theta_3\theta_4}\vline}^2\bigg{\}}.
\end{eqnarray}
The values $k$,$m$ and $l$ take the values $\{1,2,3\}$ for the
bosonic lattice states, in correspondence with the generators
$g\sim 1,f\sim 2$ and $h\sim 3$.  The fermionic terms such as
$T_{kl}$ keep the labelling $l\in \{g,f,h\}$.

The torus amplitude clearly shows the two separately connected
parts, with the sign freedom $\epsilon$ associated with those
orbits not related to the principle ones. Discrete torsion is
obtained by taking $\epsilon=-1$.  The resulting spectral content
for cases with and without torsion are quite different for both
the closed and open partition functions.  In addition, there is
the possibility of SUSY breaking in the open sector by the
possible presence of anti-branes.

The whole construction is done in the breaking from $SO(8)$ to
$SO(2)^4$ under $T^6/\bb{Z}_2\times \bb{Z}_2\times \bb{Z}_2$
orbifold compactification. The supersymmetric characters that
result from this orbifold breaking are defined as:
\begin{eqnarray}\label{eqn:char}
\tau_{oo}&=&V_2O_2O_2O_2+O_2V_2V_2V_2-S_2S_2S_2S_2-C_2C_2C_2C_2\nn\\
\tau_{og}&=&O_2V_2O_2O_2+V_2O_2V_2V_2-C_2C_2S_2S_2-S_2S_2C_2C_2\nn\\
\tau_{oh}&=&O_2O_2O_2V_2+V_2V_2V_2O_2-C_2S_2S_2C_2-S_2C_2C_2S_2\nn\\
\tau_{of}&=&O_2O_2V_2O_2+V_2V_2O_2V_2-C_2S_2C_2S_2-S_2C_2S_2C_2\nn\\
\nn\\
\tau_{go}&=&V_2O_2S_2C_2+O_2V_2C_2S_2-S_2S_2V_2O_2-C_2C_2O_2V_2\nn\\
\tau_{gg}&=&O_2V_2S_2C_2+V_2O_2C_2S_2-S_2S_2O_2V_2-C_2C_2V_2O_2\nn\\
\tau_{gh}&=&O_2O_2S_2S_2+V_2V_2C_2C_2-C_2S_2V_2V_2-S_2C_2O_2O_2\nn\\
\tau_{gf}&=&O_2O_2C_2C_2+V_2V_2S_2S_2-S_2C_2V_2V_2-C_2S_2O_2O_2\nn\\
\nn\\
\tau_{ho}&=&V_2S_2C_2O_2+O_2C_2S_2V_2-C_2O_2V_2C_2-S_2V_2O_2S_2\nn\\
\tau_{hg}&=&O_2C_2C_2O_2+V_2S_2S_2V_2-C_2O_2O_2S_2-S_2V_2V_2C_2\nn\\
\tau_{hh}&=&O_2S_2C_2V_2+V_2C_2S_2O_2-S_2O_2V_2S_2-C_2V_2O_2C_2\nn\\
\tau_{hf}&=&O_2S_2S_2O_2+V_2C_2C_2V_2-C_2V_2V_2S_2-S_2O_2O_2C_2\nn\\
\nn\\
\tau_{fo}&=&V_2S_2O_2C_2+O_2C_2V_2S_2-S_2V_2S_2O_2-C_2O_2C_2V_2\nn\\
\tau_{fg}&=&O_2C_2O_2C_2+V_2S_2V_2S_2-C_2O_2S_2O_2-S_2V_2C_2V_2\nn\\
\tau_{fh}&=&O_2S_2O_2S_2+V_2C_2V_2C_2-C_2V_2S_2V_2-S_2O_2C_2O_2\nn\\
\tau_{ff}&=&O_2S_2V_2C_2+V_2C_2O_2S_2-C_2V_2C_2O_2-S_2O_2S_2V_2.\nn\\
\end{eqnarray}
Where one combines these into the character sums as
\begin{eqnarray}
&&T_{\gamma o}=\tau_{\gamma o}+\tau_{\gamma g}+\tau_{\gamma h
}+\tau_{\gamma f} \quad \quad
T_{\gamma g}=\tau_{\gamma o}+\tau_{\gamma g}-\tau_{\gamma h}-\tau_{\gamma f}\nn\\ \nn\\
&&T_{\gamma h}=\tau_{\gamma o}-\tau_{\gamma g}+\tau_{\gamma
h}-\tau_{\gamma f} \quad \quad T_{\gamma f}=\tau_{\gamma o}-\tau_{
\gamma g}-\tau_{\gamma h}+\tau_{\gamma f}.
\end{eqnarray}
Which for the sake of clarification, $\gamma \in \{0,1,2,3\}$
where $o\sim 0$ (the $\bb{Z}_2\times \bb{Z}_2$ identity), with the
normal relations for $g$, $f$ and $h$. In addition, where ever a
sum occurs in character sets such as $T_{kl}$, it is taken that
the condition $k \neq l$ applies.

All amplitudes are one loop expressions, as such it is easily seen
that the above separates into $NS-R$ sectors, where the $-$ sign
arises from fermion statistics. The origin of the torus is thus
\begin{eqnarray}\label{eqn:OriginTorus}
{\cal
T}_0&=&\frac{1}{8}\bigg{\{}2|T_{oo}|^2+2|T_{ok}|^2+16|T_{ko}|^2+
16|T_{kk}|^2\bigg{\}}
\nn\\ \nn\\
&=&\frac{1}{8}\bigg{\{}8\big{(}|\tau_{oo}|^2+|\tau_{og}|^2+
|\tau_{of}|^2+|\tau_{oh}|^2\big{)}+64\big{(}\ldots\big{)}\bigg{\}}
\end{eqnarray}
which, similarly to the previous $\bb{Z}_2\times \bb{Z}_2$
modulated torus has 8 fixed points from each of the three twisted
sectors, as expected.

The case considered in the previous section was the projection of
the partition function by $\bb{Z}_2\times \bb{Z}_2$. Consequently,
the counting of fixed points (multiplicity factor) for the twisted
sector in the torus, in particular the $\Lambda_{m,n+\frac{1}{2}}$
massive states, is preserved as eight after the orientifold
projection. This was realized by the factor of one eighth from the
projection which includes that of the orientifold projection. In
the $\bb{Z}_2\times \bb{Z}_2\times \bb{Z}_2$ model, the freely
acting shift acts as an additional modulating group outside the
$\bb{Z}_2\times \bb{Z}_2$ projection. This requires an extra
factor of one half in the trace.  As such, the $n+\frac{1}{2}$
massive states in the twisted sector of the torus have half the
degeneracy they require for consistent interpretation as states
that exist at the eight fixed points. Therefore, the Klein must
also add an equal number of states to compensate the half from the
shift projection
\begin{eqnarray}
\frac{1}{2}\big(8_{\cal T}\Lambda_{m,n+\frac{1}{2}}+8_{\cal
K}W_{n+\frac{1}{2}}\big).
\end{eqnarray}
with eight from the torus $8_{\cal T}$, and eight from the Klein
$8_{\cal K}$.  However, the naive insertion of such a
$W_{n+\frac{1}{2}}$ term leads to inconsistent factorization in
the transverse Klein amplitude.  This inconsistency arises due to
the $S$ transformation mapping these $W_{n+\frac{1}{2}}$ states to
$(-1)^mP$.  In this case, the $O5_l-O5_k$ couplings would have a
factor of two.  A similar phenomenon arises in a six dimensional
example in \cite{CAAS}.  Here the authors consider $T^4/\bb{Z}_2$
with an unconventional orientifold projection $\xi\Omega$, for
some phase $\xi^2=1$. This model defines a direct Klein amplitude
\begin{eqnarray}
{\cal K}&=&\frac{1}{4}\bigg[(Q_o+Q_v)\bigg(\sum_m (-1)^m
\frac{q^{(\frac{\alpha^\prime}{2})m^{\rm T} g^{-1} m
}}{\eta^4}+\sum_n (-1)^n \frac{q^{(\frac{1}{2\alpha^\prime})n^{\rm
T} g n }}{\eta^4}\bigg)\nn\\
&&+2\times(n_++n_-)(Q_s+Q_c)\bigg(\frac{\eta}{\theta_4}\bigg)^4\bigg]
\end{eqnarray}
As such, in the transverse channel amplitude, the twisted sector
cannot be derived by factorization from the untwisted states that
are now entirely massive, by virtue of a redefinition of the
orientifold projection.  This then requires equal but opposite
eigenvalue assignments to the twisted states that effectively
render the counting zero with $n_+=8$ and $n_-=-8$.

The Klein amplitude for the $T^6/\bb{Z}_2\times \bb{Z}_2\times
\bb{Z}_2$ model is then given by
\begin{eqnarray}\label{eqn:DirectKlein}
{\cal
K}=\frac{1}{16}\bigg{\{}\bigg(P^{1}P^{2}P^{3}\big{(}1+(-1)^{m_1+m_2+m_3}\big{)}+
\big{(}1+(-1)^{m_1}\big{)}P^{1}W^{2}W^{3}\nn\\
+\big{(}1+(-1)^{m_2}\big{)}W^{1}P^{2}W^{3}+\big{(}1+(-1)^{m_3}\big{)}
W^{1}W^{2}P^{3}\bigg)T_{oo}\nn\\
+2\times16\epsilon_k\bigg[P^k+\epsilon W^k+(8-8)
W^k_{n+\frac{1}{2}}\bigg]\bigg{(}\frac{\eta}{\theta_4}\bigg{)}^2
T_{ko}\bigg{\}}
\end{eqnarray}
where the parameter $\epsilon$ satisfies
\begin{eqnarray}\label{eqn:Epsilon}
\epsilon=\epsilon_1\epsilon_2\epsilon_3.
\end{eqnarray}

The measure associated with the Klein for the parameter $\tau_2$
is
\begin{eqnarray}
\int\frac{d^2\tau}{{\tau_2}^3}\quad\overrightarrow{t=2\tau_2}\quad2^2
\int\frac{d^2t}{t^3}.
\end{eqnarray}
Poisson resummation gives a factor of 2 for each $T^2$ Kaluza
Klein or winding tower lattice. There is no factorial contribution
from lattices which are acted upon by an orbifold operation as
this imposes the condition of no momentum flow through the
orbifold plane. The resulting transverse Klein amplitude is then
\begin{eqnarray}
\tilde{{\cal K}}=
\frac{2^5}{16}\bigg{\{}\bigg(v_1v_2v_2(W^1_eW^2_eW^3_e+W^1_oW^2_oW^3_o)
+\frac{v_k}{2v_lv_m}W^kP^l_eP^m_e\bigg)T_{oo}
\nn\\
\nn\\+2\epsilon_k\bigg[v_kW^k_e+\epsilon\frac{P^k_e}{v_k}\bigg]{\biggl(\frac{2\eta}{\theta_2}\biggr)}^2T_{ok}\bigg{\}}
\end{eqnarray}
The usual symmetrized summation convention is used for $k$,$l$ and
$m$. The transverse Klein amplitude at the origin is
\begin{eqnarray}\label{eqn:KleinZeroModes}
\tilde{{\cal
K}}_o=\frac{2^5}{16}\bigg{\{}\bigg(v_1v_2v_2+\frac{v_k}{2v_lv_m}\bigg)
T_{oo}+2\epsilon_k\bigg(v_k+\epsilon\frac{1}{v_k}\bigg)T_{ok}\bigg{\}}
\end{eqnarray}
which has an expanded form
\begin{eqnarray}
\tilde{{\cal K}_o}=\frac{2^5}{16}&&\bigg{\{}{\bigg{(}
\sqrt{v_1v_2v_3}+\epsilon_1\sqrt{\frac{v_1}{v_2v_3}}+
\epsilon_2\sqrt{\frac{v_2}{v_1v_3}}+
\epsilon_3\sqrt{\frac{v_3}{v_1v_2}}\bigg{)}}^2\tau_{oo}\nn\\
&&+{\bigg{(}\sqrt{v_1v_2v_3}+\epsilon_1\sqrt{\frac{v_1}{v_2v_3}}-
\epsilon_2\sqrt{\frac{v_2}{v_1v_3}}-
\epsilon_3\sqrt{\frac{v_3}{v_1v_2}}\bigg{)}}^2\tau_{og}\nn\\
&&+{\bigg{(}\sqrt{v_1v_2v_3}-\epsilon_1\sqrt{\frac{v_1}{v_2v_3}}+
\epsilon_2\sqrt{\frac{v_2}{v_1v_3}}-
\epsilon_3\sqrt{\frac{v_3}{v_1v_2}}\bigg{)}}^2\tau_{of}\nn\\
&&+{\bigg{(}\sqrt{v_1v_2v_3}-\epsilon_1\sqrt{\frac{v_1}{v_2v_3}}-
\epsilon_2\sqrt{\frac{v_2}{v_1v_3}}+
\epsilon_3\sqrt{\frac{v_3}{v_1v_2}}\bigg{)}}^2\tau_{oh}\bigg{\}}.\nn\\
\end{eqnarray}
In the above expression, it is seen that the charges for the
orientifold planes can be changed in accordance to particular
model classes of the parameter (\ref{eqn:Epsilon}).

The annulus should contain $D9$ and $D5$ branes. Where in the
transverse channel, the closed string propagating between two $D9$
branes should have no total momentum flow through the boundaries.
One then has $p_L=-p_R$ which confines only winding modes to be
nonzero. Similarly for $D5$ branes which will have one lattice
with a winding tower and two with Kaluza Klein towers. The states
that flow in the torus must also flow in the transverse annulus,
so one must build on torus states using corresponding $D5$ and
$D9$ brane lattice terms.

It is appropriate for the supersymmetric character sets $T_{nm}$
to appear in the combinations
\begin{eqnarray}\label{eqn:charsusybreak}
\tilde{T}_{nm}^{(\epsilon_i)}=T_{nm}^{NS}-\epsilon_i T_{nm}^{R}.
\end{eqnarray}
Where the choice of sign can signal brane SUSY breaking. Strings
that couple to brane antibrane pairs provide character sets that
now differ from the usual supersymmetric ones (\ref{eqn:char}).
Under $S$ transformation, characters of the form
$\tilde{T}_{nm}^{(-1)}$ are the same as in (\ref{eqn:char}) except
for the changes of $O_2\leftrightarrow V_2$ and
$S_2\leftrightarrow C_2$ in the last three factors. The characters
that correspond to $\tilde{T}_{nm}^{(+1)}$ are simply denoted
$T_{nm}$.
\begin{eqnarray}\label{eqn:NonSUSYChar}
\tau_{oo}^{(-1)}&=&O_2O_2O_2O_2+V_2V_2V_2V_2-C_2S_2S_2S_2-S_2C_2C_2C_2 \nn\\
\tau_{og}^{(-1)}&=&V_2V_2O_2O_2+O_2O_2V_2V_2-S_2C_2S_2S_2-C_2S_2C_2C_2 \nn\\
\tau_{oh}^{(-1)}&=&V_2O_2O_2V_2+O_2V_2V_2O_2-S_2S_2S_2C_2-C_2C_2C_2S_2 \nn\\
\tau_{of}^{(-1)}&=&V_2O_2V_2O_2+O_2V_2O_2V_2-S_2S_2C_2S_2-C_2C_2S_2C_2 \nn\\
\nn\\
\tau_{go}^{(-1)}&=&O_2O_2S_2C_2+V_2V_2C_2S_2-C_2S_2V_2O_2-S_2C_2O_2V_2 \nn\\
\tau_{gg}^{(-1)}&=&V_2V_2S_2C_2+O_2O_2C_2S_2-C_2S_2O_2V_2-S_2C_2V_2O_2 \nn\\
\tau_{gh}^{(-1)}&=&V_2O_2S_2S_2+O_2V_2C_2C_2-S_2S_2V_2V_2-C_2C_2O_2O_2 \nn\\
\tau_{gf}^{(-1)}&=&V_2O_2C_2C_2+O_2V_2S_2S_2-C_2C_2V_2V_2-S_2S_2O_2O_2 \nn\\
\nn\\
\tau_{ho}^{(-1)}&=&O_2S_2C_2O_2+V_2C_2S_2V_2-S_2O_2V_2C_2-C_2V_2O_2S_2 \nn\\
\tau_{hg}^{(-1)}&=&V_2C_2C_2O_2+O_2S_2S_2V_2-S_2O_2O_2S_2-C_2V_2V_2C_2 \nn\\
\tau_{hh}^{(-1)}&=&V_2S_2C_2V_2+O_2C_2S_2O_2-C_2O_2V_2S_2-S_2V_2O_2C_2 \nn\\
\tau_{hf}^{(-1)}&=&V_2S_2S_2O_2+O_2C_2C_2V_2-S_2V_2V_2S_2-C_2O_2O_2C_2 \nn\\
\nn\\
\tau_{fo}^{(-1)}&=&O_2S_2O_2C_2+V_2C_2V_2S_2-C_2V_2S_2O_2-S_2O_2C_2V_2 \nn\\
\tau_{fg}^{(-1)}&=&V_2C_2O_2C_2+O_2S_2V_2S_2-S_2O_2S_2O_2-C_2V_2C_2V_2 \nn\\
\tau_{fh}^{(-1)}&=&V_2S_2O_2S_2+O_2C_2V_2C_2-S_2V_2S_2V_2-C_2O_2C_2O_2 \nn\\
\tau_{ff}^{(-1)}&=&V_2S_2V_2C_2+O_2C_2O_2S_2-S_2V_2C_2O_2-C_2O_2S_2V_2. \nn\\
\end{eqnarray}
%
%


\subsection{Open Descendants}


The addition of terms in the torus that define contributions from
orbits that lie outside the connection of $S$ and $T$ transforms
comes with the sign freedom $\epsilon=\pm 1$. Such differences
are already evident in the closed amplitudes, as will be shown,
different choice lead to quite distinct open amplitudes with very
different phenomenological characteristics.

Considering only the $g$-twisted sector, since the others will
follow the same principles, the torus at massless level has
contributions from
\begin{eqnarray}\label{eqn:TorusTwistedOrigin}
{\cal T}^g_o
&=&4(\epsilon+1)\big(|\tau_{go}|^2+|\tau_{go}|^2+|\tau_{gg}|^2+|\tau_{gf}|^2+|\tau_{gh}|^2\big)\nn\\
&&+2(\epsilon-1)\big(\tau_{go}\bar{\tau}_{gg}+\tau_{gg}\bar{\tau}_{go}\big).\nn\\
\end{eqnarray}
In the transverse channel, the annulus is defined as closed string
states propagating between boundaries that in this case are either
$D9$ or $D5$ branes.  Since world sheet time is now in a direction
orthogonal to the boundaries, one has states of the form $<{\rm
Final}|$ and $|{\rm Initial}>$ which are CPT conjugates. By
reference to the supersymmetric $SO(2)^4$ characters
(\ref{eqn:char}), it is seen that the second line of
(\ref{eqn:TorusTwistedOrigin}) contains such conjugate pairs.  So
for $\epsilon=+1$, there are no twisted states that propagate in
the transverse channel, for the case $\epsilon=-1$, such states
are allowed.


\subsection{Models Without Discrete Torsion $(\epsilon=+1)$}


The subclass of models is generated by $\epsilon=(+,+,+)$, where
$\epsilon_k=+1$, and $(+,-,-)$ with two additional permutations
$(-,+,-)$ and $(-,-,+)$.  As has been shown in
(\ref{eqn:charsusybreak}), the presence of any $\epsilon_k=-1$
breaks supersymmetry.

The transverse annulus amplitude is defined by
\begin{eqnarray}\label{eqn:TransverseAnnulusWT}
\tilde{{\cal
A}}=&\frac{2^{-5}}{16}&\bigg{\{}\bigg{(}N_o^2v_1v_2v_3(W^1W^2W^3+
W^1_{n+\frac{1}{2}}W^2_{n+\frac{1}{2}}W^3_{n+\frac{1}{2}})\nn\\
&&+D^2_{ko}\frac{v_k}{2v_lv_s}W^kP^lP^s\frac{\big{(}1+(-1)^{m_l+m_s}\big{)}}{2}
\bigg{)}T_{oo}\nn\\
&&+2D_{ko}Nv_kW^k\bigg{(}\frac{2\eta}{\theta_2}\bigg{)}^2\tilde{T}_{ok}^{(\epsilon_k)}\nn\\
&&+D_{ko}D_{lo}\frac{1}{v_s}\frac{P^s(1+(-1)^s)}{2}\bigg{(}\frac{2\eta}{\theta_2}\bigg{)}^2\tilde{T}_{os}^{(\epsilon_k
\epsilon_l )}\nn\\
\end{eqnarray}
where the construction follows from that done in the (shift)
orientifold case, with the exception of the $D5_k-D5_l$
interactions. These follow from the boundary conditions of the
branes and the towers that are allowed by the torus.

The numerical coefficients are to begin with undetermined, and are
constrained by the requiring that (\ref{eqn:AnnulusOrigin1}) is
obeyed. The origin of the lattice towers shows the perfect square
structure as
\begin{eqnarray}\label{eqn:AnnulusOrigin1}
\tilde{{\cal
A}}_{o}&=&\frac{2^{-5}}{16}\bigg{\{}{\bigg{(}N_o\sqrt{v_1v_2v_3}+
D_{go}\sqrt{\frac{v_1}{v_2v_3}}+D_{fo}\sqrt{\frac{v_2}{v_lv_3}}+
D_{ho}\sqrt{\frac{v_3}{v_1v_2}}\bigg{)}}^2\tau_{oo}^{NS}\nn\\ \nn\\
&&-{\bigg{(}N_o\sqrt{v_1v_2v_3}+\epsilon_1D_{go}\sqrt{\frac{v_1}{v_2v_3}}
+\epsilon_2D_{fo}\sqrt{\frac{v_2}{v_1v_3}}+\epsilon_3D_{ho}
\sqrt{\frac{v_3}{v_1v_2}}\bigg{)}}^2\tau_{oo}^{R}\nn\\ \nn\\
&&+{\bigg{(}N_o\sqrt{v_1v_2v_3}+D_{go}\sqrt{\frac{v_1}{v_2v_3}}
-D_{fo}\sqrt{\frac{v_2}{v_1v_3}}-D_{ho}\sqrt{\frac{v_3}{v_1v_2}}
\bigg{)}}^2\tau_{og}^{NS}\nn\\ \nn\\
&&-{\bigg{(}N_o\sqrt{v_1v_2v_3}+\epsilon_1D_{go}\sqrt{\frac{v_1}{v_2v_3}}
-\epsilon_2D_{fo}\sqrt{\frac{v_2}{v_1v_3}}-\epsilon_3D_{ho}
\sqrt{\frac{v_3}{v_1v_2}}\bigg{)}}^2\tau_{og}^{R}\nn\\ \nn\\
&&+{\bigg{(}N_o\sqrt{v_1v_2v_3}-D_{go}\sqrt{\frac{v_1}{v_2v_3}}
+D_{fo}\sqrt{\frac{v_2}{v_1v_3}}-D_{ho}\sqrt{\frac{v_3}{v_1v_2}}
\bigg{)}}^2\tau_{of}^{NS}\nn\\ \nn\\
&&-{\bigg{(}N_o\sqrt{v_1v_2v_3}-\epsilon_1D_{go}\sqrt{\frac{v_1}{v_2v_3}}
+\epsilon_2D_{fo}\sqrt{\frac{v_2}{v_1v_3}}-\epsilon_3D_{ho}
\sqrt{\frac{v_3}{v_1v_2}}\bigg{)}}^2\tau_{of}^{R}\nn\\ \nn\\
&&+{\bigg{(}N_o\sqrt{v_1v_2v_3}-D_{go}\sqrt{\frac{v_1}{v_2v_3}}-
D_{fo}\sqrt{\frac{v_2}{v_1v_3}}+D_{ho}\sqrt{\frac{v_3}{v_1v_2}}
\bigg{)}}^2\tau_{oh}^{NS}\nn\\ \nn\\
&&-{\bigg{(}N_o\sqrt{v_1v_2v_3}-\epsilon_1D_{go}\sqrt{\frac{v_1}{v_2v_3}}
-\epsilon_2D_{fo}\sqrt{\frac{v_2}{v_1v_3}}+\epsilon_3D_{ho}\sqrt{\frac{v_3}
{v_1v_2}}\bigg{)}}^2\tau_{oh}^{R}\bigg{\}}.\nn\\
\end{eqnarray}
An $S$ transform shows the direct channel amplitude to be
\begin{eqnarray}\label{eqn:DirectA}
{\cal A}&=&\frac{1}{16} \bigg\{\bigg(N^2P_1P_2P_3\big(1+(-1)^{m_1+m_2+m_3}\big)\nn\\
&&+\frac{1}{4}D^2_{ko}P^k\big(W^lW^s+W^l_{n+\frac{1}{2}}W^s_{n+\frac{1}{2}}\big)\bigg)T_{oo}\nn\\
&&+2D_{ko}NP^k\bigg{(}\frac{\eta}{\theta_4}\bigg{)}^2\tilde{T}_{ko}^{(\epsilon_k)}\nn\\
&&\frac{1}{2}D_{ko}D_{lo}\big(W^s+W^s_{n+\frac{1}{2}}\big)\bigg{(}\frac{\eta}{\theta_4}\bigg{)}^2\tilde{T}_{so}^{(\epsilon_k
\epsilon_l )}\bigg\}\nn\\
\end{eqnarray}
Where it can be seen from (\ref{eqn:char}) that the following
character sets transform in the following manner under $S$:
\begin{eqnarray}
T_{om}\rightarrow T_{mo}
\end{eqnarray}
which is relevant to the $\epsilon=+1$ models, and
\begin{eqnarray}
T_{mm}\rightarrow-T_{mm},\quad\quad T_{kl}\rightarrow
i(-1)^{k+l}T_{kl},\quad\quad
\end{eqnarray}
for those characters that appear for the cases of $\epsilon=-1$.
This can be seen simply by acting with the operator $S$ on the
characters (\ref{eqn:char}), which has the form
\begin{eqnarray}
S_{2n}=\frac{1}{2}\left(\matrix{ 1 & 1 & 1 & 1 \cr 1 & 1& -1 & -1
\cr 1 & -1 & i^{-n} & -i^{-n} \cr 1 & -1 & -i^{-n} & i^{-n} \cr
}\right),
\end{eqnarray}
and acts on the transverse of the vector
$(O_{2n},V_{2n},S_{2n},C_{2n})$ for the characters of $SO(2n)$.

Equation (\ref{eqn:TransverseAnnulusWT}) highlights the
arrangement of the $D5$ branes which are shown in figure
\ref{aZ2xZ2xZ21}, with arrows indicating the interchanges of the
various $\bb{Z}_2\times \bb{Z}_2\times \bb{Z}_2$ generators.  The
diagram shows the placements of $D5_{go}$, $D5_{fo}$ and $D5_{ho}$
branes within the annulus amplitude according to the coordinates
they wrap, which can be seen in the amplitude as the
correspondences of the wrapping $D_{go}\sim$45, $D_{fo}\sim$67 and
$D_{ho}\sim$89. Figure \ref{aZ2xZ2xZ21} highlights the generic
feature of this freely acting shift on the relatively simple
geometry (in comparison to those considered in \cite{AADS}, which
are freely acting orbifolds with non-freely acting winding and or
momentum shifts).
%
%
\begin{figure}[!ht]
\centerline{\epsfxsize 2.4 truein \epsfbox {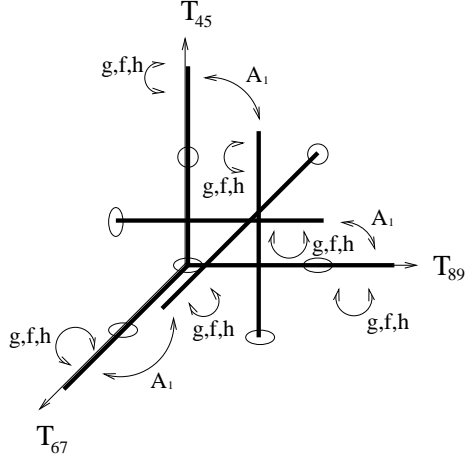}}
\caption{$D5_{ko}$ configurations} \label{aZ2xZ2xZ21}
\end{figure}
%
%

In constructing the Mobius it will be necessary to perform $P$
transforms on the amplitude components in order to gain the direct
channel equation.  Formally, the $P$ operator is a combination of
the already understood $S$ and $T$ transforms as
\begin{eqnarray}
P=TST^2S\nn
\end{eqnarray}
and so acts on the measure as
\begin{eqnarray}\label{eqn:mobmeas}
P:\frac{1}{2}+i\frac{\tau_2}{2}\quad\rightarrow \quad
\frac{1}{2}+\frac{i}{2{\tau_2}}
\end{eqnarray}
Combinations of the $S$ and $T$ operators satisfy
\begin{eqnarray}
S^2=(ST)^3=C\quad\Rightarrow\quad P^2=C
\end{eqnarray}
where $C$ is the charge conjugation matrix, which in these cases,
is simply the identity. It is seen that this operation acts on the
Mobius lattice modes as an $S$ transform on a Klein lattice, as
shown in table \ref{tab:Stransforms}. This then implies an action
on the characters as
\begin{eqnarray}\label{eqn:Pmatrix}
P=\left(\matrix{ c & s & 0 & 0 \cr s & -c & 0 & 0 \cr 0 & 0 & \chi
c & i\chi s \cr 0 & 0 & i\chi s & \chi c }\right)
\end{eqnarray}
for $s=sin(n\pi/4)$, $c=cos(n\pi/4)$ and
$\chi=e^{-i\frac{n\pi}{4}}$, for an $SO(2n)$ breaking.
\begin{table}
\begin{center}
\begin{tabular}{|c|c|}\hline
Plane diagrams & Volumes \\ \hline $D9-D9$, $D9-O9$,
$O9-O9$ & $v_1v_2v_3$ \\
$D5_k-D5_k$, $D5_k-O5_k$, $O5_k-O5_k$ & $\frac{v_k}{v_lv_m}$ \\
$D5_k-D5_l$, $D5_k-O5_l$, $O5_k-O5_l$ & $\frac{1}{v_m}$ \\
$D9-D5_k$, $D9-O5_k$, $D5_k-O9$ & $v_k$ \\ \hline \hline
$\tilde{\cal A}$ and $\tilde{\cal K}$ Plane Diagrams & Lattice
Couplings
\\\hline
$D9-D9$ & $W^1W^2W^3+W^1_{n+\frac{1}{2}}W^2_{n+\frac{1}{2}}
W^3_{n+\frac{1}{2}}$
\\
$D5_k-D5_k$ & $W^kP^lP^m\big(1+(-1)^{m_l+m_m}\big)$ \\
$O9-O9$ & $W^1_eW^2_eW^3_e+W^1_oW^2_oW^3_o$ \\
$O5_k-O5_k$ & $W^kP^l_eP^m_e$ \\ \hline \hline $\tilde{\cal M}$
plane diagrams & Lattice Couplings
\\\hline
$D9-O9$ & $W^1_eW^2_eW^3_e+W^1_oW^2_oW^3_o$ \\
$D9-O5_k$ & $W^k$ \\
$D5_k-O9$ & $W^k_e$ \\
$D5_k-O5_k$ & $W^k_eP^l_eP^m_e+(-1)^{m_l+m_m}W^k_oP^l_eP^m_e$ \\
$D5_k-O5_l$ & $P^m_e$ \\\hline
\end{tabular}
\end{center}
\caption{Lattice restrictions}\label{tab:cc}
\end{table}

Having fixed the relevant factors in the annulus, the Mobius can
now be constructed. The Mobius should symmetrize the Klein and the
annulus in the transverse channel, and must give a proper particle
interpretation with the annulus in the direct channel. Since the
Mobius in the transverse channel is a closed string propagating
between a brane and an $O$ plane, it is necessary to understand
the constraints on the Kaluza Klein and winding terms placed by
the brane and $O$-plane diagrams.  Table \ref{tab:cc} shows the
constrained lattice terms as read off of $\tilde{{\cal A}}$ and
$\tilde{{\cal K}}$. Taking the cross couplings of the various
$\tilde{\tau}_{ok}$ from the origin states of $\tilde{{\cal A}}_o$
and $\tilde{{\cal K}}_o$, the Mobius origin reads
\begin{eqnarray}\label{eqn:mobzm}
\tilde{{\cal
M}}_o&=&\pm\frac{1}{8}\bigg{\{}N_ov_1v_2v_3\hat{T}_{oo}+\epsilon_k
D_{ko}\frac{v_k}{2v_lv_m}\hat{\tilde{T}}_{oo}^{(\epsilon_k)}\nn\\
&&+\epsilon_kN_ov_k\hat{\tilde{T}}_{ok}+D_{ko}v_k
\hat{\tilde{T}}_{ok}^{(\epsilon_k)}+\epsilon_mD_{lo}\frac{1}{v_k}
\hat{\tilde{T}}_{ok}^{(\epsilon_l)}\bigg{\}}
\end{eqnarray}
The hatted characters signify the Mobius measure as defined in
(\ref{eqn:mobmeas}).  From here, the Kaluza Klein or winding
towers must be built by taking the common states from
$\tilde{A}_o$ and $\tilde{{\cal K}}_o$.  Such towers are fully
illustrated in table \ref{tab:cc}. In the direct annulus, there
are only integer lattice modes present on the $D9$-$D9$ coupling,
so the resulting term is thus
\begin{eqnarray}
\tilde{\cal
M}_{O9-D9}=\pm\frac{N_o}{8}v_1v_2v_3(W^1_eW^2_eW^3_e+W^1_oW^2_oW^3_o)
\hat{T}_{oo}
\end{eqnarray}
The coupling of the $D_{ko}$ and the $O5_k$ presents a subtlety.
Since the shift $A_1$ acts in one coordinate of each of the
internal directions, the direct channel Klein $O5-O5$ couplings
each have a projected momentum lattice as $(1+(-1)^{m_k})P^k$.
This leads to the common lattice modes between the transverse
annulus and Klein amplitudes to have all winding modes.  This
would lead to inconsistent symmetrization with the direct channel
annulus, since the corresponding lattice in the annulus is $P^k$
and that of the direct Mobius would be $2P^k_e$.  This can be
rectified by splitting the winding mode lattice to $W_e+W_o$ and
introducing a phase with the momentum modes that exist with the
odd winding modes, so that
\begin{eqnarray}
\tilde{\cal
M}_{D5_k-O5_k}=\pm\frac{1}{8}\epsilon_kD_{ko}\frac{v_k}{2v_lv_s}
\big(W^k_eP^l_eP^m_e+(-1)^{m_l+m_s}W^k_oP^l_eP^s_e\big)
\end{eqnarray}
This then allows the proper symmetrization between the direct
annulus and Mobius for integer lattice modes.  This type of
modification is also seen in the (shift) orbifold model considered
previously.

This now exhausts the Mobius terms due to the form of the Klein,
which is free of twisted terms, so one finds
\begin{eqnarray}
\tilde{{\cal
M}}&=&-\frac{1}{8}\bigg{\{}N_ov_1v_2v_3(W^1_eW^2_eW^3_e+W^1_oW^2_oW^3_o)
\hat{T}_{oo}\nn\\
&&+\epsilon_kD_{ko}\frac{v_k}{2v_lv_s}\big(W^k_eP^l_eP^s_e+(-1)^{m_l+m_s}
W^k_oP^l_eP^s_e\big)\hat{\tilde{T}}_{oo}^{(\epsilon_k)}\nn\\
&&+\big(\epsilon_kN_ov_kW^k\hat{\tilde{T}}_{ok}+D_{ko}v_kW^k_e
\hat{\tilde{T}}_{ok}^{(\epsilon_k)}\big){\bigg(\frac{2\hat{\eta}}
{\hat{\theta}_2}\bigg)}^2\nn\\
&&+\epsilon_mD_{lo}\frac{P^k_e}{v_k}\hat{\tilde{T}}_{ok}^{(\epsilon_l)}
{\bigg(\frac{2\hat{\eta}}{\hat{\theta}_2}\bigg)}^2\bigg{\}}
\end{eqnarray}

The Mobius origin (\ref{eqn:mobzm}) yields different charges for
the brane-$O$-plane couplings. The sign ambiguity from the
coupling constants is restricted to $(-)$ in all diagrams, as will
be apparent for tadpole cancellation.  All diagrams must also have
the same sign to yield a consistent Mobius origin structure
(\ref{eqn:mobzm}).

The corresponding direct channel is obtained by $P$
transformation. It is noted that while $P$ has non trivial effect
of the lattice modes, it leaves the characters unchanged with the
exception of a sign change for the orbifold sector. This can be
seen by the representation defined in (\ref{eqn:Pmatrix}).  As
such
\begin{eqnarray}\label{eqn:DirectM}
{\cal
M}&=&-\frac{1}{16}\bigg{\{}N_oP^1P^2P^3\big(1+(-1)^{m_1+m_2+m_3}\big)
\hat{T}_{oo}\nn\\
&&+\epsilon_k\frac{1}{2}D_{ko}\big(P^kW^lW^s+(-1)^{m_k}P^k
W^l_{n+\frac{1}{2}}W^s_{n+\frac{1}{2}}\big)\hat{T}_{oo}^{(\epsilon_k)}\nn\\
&&-\big(2\epsilon_kN_oP^k_e\hat{T}_{ok}+D_{ko}P^k
\hat{T}_{ok}^{(\epsilon_k)}\big){\bigg(\frac{2\hat{\eta}}{\hat{\theta}_2}
\bigg)}^2\nn\\
&&-\epsilon_mD_{lo}W^k\hat{T}_{ok}^{(\epsilon_l)}{\bigg(\frac{2\hat{\eta}}
{\hat{\theta}_2}\bigg)}^2\bigg{\}}
\end{eqnarray}

By virtue of (\ref{eqn:charsusybreak}), it would seem that there
are tachyonic modes present here with the presence of
$V_2O_2O_2O_2\rightarrow O_2O_2O_2O_2$ (by reference to
(\ref{eqn:NonSUSYChar})). However, the $P$ transformation acting
on the characters is structured differently form the usual $S$
transformation, consequently such a mass change odes not exist in
the Mobius. The terms in the direct channel Mobius amplitude are
thus free of tachyonic states. The direct annulus has no terms of
the form $T_{ok}^{(-1)}$ or $T_{oo}^{(-1)}$, and so the model is
tachyon free.  This generically follows from the parent
$\bb{Z}_2\times \bb{Z}_2$ model, and thus is true for any further
shift modulation of it.

The tadpole conditions for the $D9$ branes are
\begin{eqnarray}\label{eqn:Ntadpole}
\frac{2^5}{16}+\frac{2^{-5}}{16}N_o^2-\frac{N_o}{8}=0~\Rightarrow~N_o=32.
\end{eqnarray}
It is seen that the tadpole conditions in the $NS$ and $R$ sector
can not lead to mutual cancellation of tadpoles for cases other
than $(+,+,+)$. The tadpole for $N_o$ (\ref{eqn:Ntadpole}), is
unaffected by this. However, allowing the cancellation of the $R$
sector forces a tree level dilaton tadpole correlated with a
potential for the geometric moduli to be created.  This has an
interpretation of increased vacuum energy.  The tadpoles arising
from the Ramond sector must be satisfied in order to suppress
anomalies.  From the amplitudes, one finds
\begin{eqnarray}\label{eqn:Dtadpole}
D_{ko}^{(NS)}=\epsilon_k32{\rm ,}~D_{ko}^{(R)}=32.
\end{eqnarray}


\subsection{Model Classes of $\epsilon=+1$}


The direct channel amplitudes defined by eqns. (\ref{eqn:DirectA})
and (\ref{eqn:DirectM}) require a rescaling of $N\rightarrow 2N$
and $D_{ko}\rightarrow 4D_{ko}$ to be consistent.  As such, one
now has
\begin{eqnarray}\label{eqn:AnnulusRescaled}
{\cal A} &=&\frac{1}{4}\bigg\{\bigg(n^2P_1P_2P_3\big(1+(-1)^{m_1+m_2+m_3}\big)\nn\\
&&+d^2_{ko}P^k\big(W^lW^s+W^l_{n+\frac{1}{2}}W^s_{n+\frac{1}{2}}\big)\bigg)T_{oo}\nn\\
&&+4d_{ko}nP^k\bigg{(}\frac{\eta}{\theta_4}\bigg{)}^2T_{ko}^{(\epsilon_k)}\nn\\
&&+4d_{ko}d_{lo}\frac{\big(W^s+W^s_{n+\frac{1}{2}}\big)}{2}\bigg{(}\frac{\eta}{\theta_4}\bigg{)}^2T_{mo}^{(\epsilon_k\epsilon_l)}\bigg\}
\end{eqnarray}
and
\begin{eqnarray}\label{eqn:MobiusRescaled}
{\cal
M}&=&-\frac{1}{8}\bigg{\{}nP^1P^2P^3\big(1+(-1)^{m_1+m_2+m_3}\big)
\hat{T}_{oo}\nn\\
&&+\epsilon_kd_{ko}\big(P^kW^lW^s+(-1)^{m_k}P^k
W^l_{n+\frac{1}{2}}W^s_{n+\frac{1}{2}}\big)\hat{T}_{oo}^{(\epsilon_k)}\nn\\
&&-\big(2\epsilon_knP^k_e\hat{T}_{ok}+2d_{ko}P^k
\hat{T}_{ok}^{(\epsilon_k)}\big){\bigg(\frac{2\hat{\eta}}{\hat{\theta}_2}
\bigg)}^2\nn\\
&&-\epsilon_m2d_{lo}W^k\hat{T}_{ok}^{(\epsilon_l)}{\bigg(\frac{2\hat{\eta}}
{\hat{\theta}_2}\bigg)}^2\bigg{\}}.
\end{eqnarray}
Here, it appears that the massive modes, in particular the
$(-1)^{m_k}P^kW^l_{n+\frac{1}{2}}W^s_{n+\frac{1}{2}}$ towers.  The
annulus and Mobius are required to symmetrize modulo two.  In this
case, one has a multiplicity of the common states in the Mobius
and annulus as
\begin{eqnarray}
2^3\times\bigg(\frac{d_{ko}^2}{4}-\frac{d_{ko}}{8}\bigg)=3\frac{d_{ko}(d_{ko}-1)}{2}+\frac{d_{ko}(d_{ko}+1)}{2}
\end{eqnarray}
where the multiplicity of $2^3$ comes from the interchange of the
indices $l$ and $s$ and the degeneracy of massive states under an
orbifold element as $\alpha:n+\frac{1}{2}\rightarrow
-n-\frac{1}{2}$ for $\alpha\in \bb{Z}_2$.  So one finds that group
interpretation is preserved as the decomposition into three
orthogonal copies and one simplectic.

Firstly, we discuss the simplest and supersymmetric case of
$(+,+,+)$.  The massless spectra of (\ref{eqn:AnnulusRescaled})
and (\ref{eqn:MobiusRescaled}) is
\begin{eqnarray}\label{eqn:AnnulusMobiusOrigin1}
{\cal A}_o+{\cal M}_o
=&&\bigg[\frac{n(n+1)}{2}+\frac{d_{go}(d_{go}+1)}{2}\nn\\
&&+\frac{d_{fo}(d_{fo}+1)}{2}+\frac{d_{ho}(d_{ho}+1)}{2}\bigg]\tau_{oo}\nn\\
&&\bigg[\frac{n(n-1)}{2}+\frac{d_{go}(d_{go}-1)}{2}\nn\\
&&+\frac{d_{fo}(d_{fo}-1)}{2}+\frac{d_{ho}(d_{ho}-1)}{2}\bigg](\tau_{og}+\tau_{of}+\tau_{oh})\nn\\
&&+(nd_{go}+d_{fo}d_{ho})(\tau_{go}+\tau_{gg}+\tau_{gf}+\tau_{gh})\nn\\
&&+(nd_{fo}+d_{go}d_{ho})(\tau_{fo}+\tau_{fg}+\tau_{ff}+\tau_{fh})\nn\\
&&+(nd_{ho}+d_{go}d_{fo})(\tau_{ho}+\tau_{hg}+\tau_{hf}+\tau_{hh}).\nn\\
\end{eqnarray}
The vector multiplet, contained in $\tau_{oo}$, combined with the
tadpole conditions (\ref{eqn:Ntadpole}) and (\ref{eqn:Dtadpole})
with the rescaling $N=2n$ and $D_{ko}=4d_{ko}$ shows the gauge
group to be $USp(16)_9\times USp(8)_{5_{\{1,2,3\}}}$.  Where the
suffixes refer to the groups of the $D9$ and three copies of $D5$.
{}From (\ref{eqn:char}), one can see that at ${\cal N}=1$, chiral
multiplets arise in the untwisted sector from $\tau_{ok}$ and from
the twisted sector in $\tau_{gf}$, $\tau_{hg}$ and $\tau_{fg}$.
This model therefore has chiral multiplets in the representations
described in table \ref{tab:ChiralReps1}
\begin{table}[!ht]
\begin{center}
\begin{tabular}{|l|l|l|}\hline
Sector & & Reps. in $(D9,D5_k)$ \\
\hline Twisted & $\tau_{gf}$ & $(16,8_1)+(1,8_2\oplus 8_3)$
\\
& $\tau_{hg}$ & $(16,8_3)+(1,8_1\oplus 8_2)$
\\
& $\tau_{fg}$ & $(16,8_2)+(1,8_1\oplus 8_3)$
\\\hline
Untwisted & $\tau_{ok}$ & $(120,28_1\oplus 28_3\oplus 28_3)$
\\\hline
\end{tabular}
\end{center}
\caption{Chiral multiplet representations for
$\epsilon=(+,+,+)$}\label{tab:ChiralReps1}
\end{table}

The remaining cases break supersymmetry for states coupling to
$D5$ branes that are aligned with the directions that satisfy
$\epsilon_k=-1$.  In fact, such branes that exist in such
directions are interpreted as antibranes.

For $(+,-,-)$, one has low lying spectrum
\begin{eqnarray}\label{eqn:AnnulusMobiusOrigin2}
{\cal A}_o+{\cal M}_o
=&&\bigg[\frac{n(n-1)}{2}+\frac{d_{go}(d_{go}-1)}{2}\bigg](\tau_{oo}+\tau_{of}+\tau_{oh})\nn\\
&&\bigg[\frac{n(n+1)}{2}+\frac{d_{go}(d_{go}+1)}{2}\bigg]\tau_{og}\nn\\
&&+\bigg[\frac{d_{fo}(d_{fo}-1)}{2}+\frac{d_{ho}(d_{ho}-1)}{2}\bigg]
\tau_{og}^{\rm NS}\nn\\
&&+\bigg[\frac{d_{fo}(d_{fo}+1)}{2}+\frac{d_{ho}(d_{ho}+1)}{2}\bigg]
\tau_{og}^{\rm R}\nn\\
&&+\bigg[\frac{d_{fo}(d_{fo}+1)}{2}+\frac{d_{ho}(d_{ho}+1)}{2}\bigg]
(\tau_{oo}^{\rm NS}+\tau_{of}^{\rm NS}+\tau_{oh}^{\rm NS})\nn\\
&&+\bigg[\frac{d_{fo}(d_{fo}-1)}{2}+\frac{d_{ho}(d_{ho}-1)}{2}\bigg]
(\tau_{oo}^{\rm R}+\tau_{of}^{\rm R}+\tau_{oh}^{\rm R})\nn\\
&&+(nd_{go}+d_{fo}d_{ho})(\tau_{go}+\tau_{gg}+\tau_{gf}+\tau_{gh})\nn\\
&&+(nd_{fo}+d_{go}d_{ho})(\tau_{fo}^{(-)}+\tau_{fg}^{(-)}+\tau_{ff}^{(-)}+\tau_{fh}^{(-)})\nn\\
&&+(nd_{ho}+d_{go}d_{fo})(\tau_{ho}^{(-)}+\tau_{hg}^{(-)}+\tau_{hf}^{(-)}+\tau_{hh}^{(-)})\nn\\
\end{eqnarray}
Supersymmetry is seen to be broken in this expansion in a twofold
way.  Firstly, the representations for the Neveu-Schwartz and
Ramond sectors are different.  Secondly, the presence of signs
from $\epsilon_k$ in the Ramond sector transforms differently
under $S$.  It can be seen by reference to (\ref{eqn:NonSUSYChar})
that the masses of multiplet components is now different.  In this
case, the gauge group is $SO(19)_9\times SO(8)_{5_1}\times
USp(8)_{\bar{5}_{\{2,3\}}}$.  The chiral representations are
displayed in table \ref{tab:ChiralReps2}.
\begin{table}[!ht]
\begin{center}
\begin{tabular}{|l|l|l|}\hline
Sector & & Reps. in $(D9,D5_k)$ \\
\hline Twisted & $\tau_{gf}$ & $(16,8_1)+(1,8_2\oplus 8_3)$
\\\hline
Untwisted & $\tau_{og}$ & $(136,36_1)$
\\
& $\tau_{of}$ & $(120,28_1)$
\\
& $\tau_{oh}$ & $(120,28_1)$
\\
\hline
\end{tabular}
\end{center}
\caption{Chiral multiplet representations for
$\epsilon=(+,-,-)$}\label{tab:ChiralReps2}
\end{table}
\begin{table}[!ht]
\begin{center}
\begin{tabular}{|l|l|l|}\hline
Sector & & Reps. in $(D9,D5_k)$ \\
\hline Twisted & $\tau_{fg}$ & $(16,8_2)+(1,8_1\oplus 8_3)$
\\\hline
Untwisted & $\tau_{og}$ & $(120,28_2)$
\\
& $\tau_{of}$ & $(136,36_2)$
\\
& $\tau_{oh}$ & $(120,28_2)$
\\
\hline
\end{tabular}
\end{center}
\caption{Chiral multiplet representations for
$\epsilon=(-,+,-)$}\label{tab:ChiralReps3}
\end{table}
\begin{table}[!ht]
\begin{center}
\begin{tabular}{|l|l|l|}\hline
Sector & & Reps. in $(D9,D5_k)$ \\
\hline Twisted & $\tau_{fg}$ & $(16,8_3)+(1,8_1\oplus 8_2)$
\\\hline
Untwisted & $\tau_{og}$ & $(120,28_3)$
\\
& $\tau_{of}$ & $(120,28_3)$
\\
& $\tau_{oh}$ & $(136,36_3)$
\\
\hline
\end{tabular}
\end{center}
\caption{Chiral multiplet representations for
$\epsilon=(-,-,+)$}\label{tab:ChiralReps4}
\end{table}
The remaining models behave in a similar way to the model
considered above, and for brevity, we only state their
corresponding Mobius amplitude which governs the group
representations.  For $(-,+,-)$, the Mobius is
\begin{eqnarray}
{\cal M}_o
=&&\frac{1}{2}(n+d_{fo})(-\tau_{oo}-\tau_{og}+\tau_{of}-\tau_{oh})\nn\\
&&\frac{1}{2}(d_{go}+d_{ho})(\tau_{oo}^{(-)}+\tau_{og}^{(-)}-\tau_{of}^{(-)}+\tau_{oh}^{(-)})\nn\\
\end{eqnarray}
which gauge group $SO(19)_9\times USp(8)_{\bar{5}_1}\times
SO(8)_{5_2}\times USp(8)_{\bar{5}_{3}}$ and chiral representations
displayed in table \ref{tab:ChiralReps3}.

Finally, the $(-,-,+)$ model gives rise to
\begin{eqnarray}
{\cal M}_o
=&&\frac{1}{2}(n+d_{ho})(-\tau_{oo}-\tau_{og}-\tau_{of}+\tau_{oh})\nn\\
&&\frac{1}{2}(d_{go}+d_{fo})(\tau_{oo}^{(-)}+\tau_{og}^{(-)}+\tau_{of}^{(-)}-\tau_{oh}^{(-)})\nn\\
\end{eqnarray}
with $SO(19)_9\times USp(8)_{\bar{5}_{\{1,2\}}}\times SO(8)_{5_3}$
gauge group.  Chiral representations are displayed in table
\ref{tab:ChiralReps4}.

The twisted sectors of the last two cases have broken
supersymmetry on branes that are aligned with the $\epsilon_k=-1$
directions.


\subsection{Models With Discrete Torsion $(\epsilon=-1)$}


The oriented open sector for this class of models is far more rich
than those without discrete torsion.  By reference to
(\ref{eqn:TorusTwistedOrigin}), one has the existence of left
moving states coupled to their corresponding CPT conjugates for
$\epsilon=-1$.  In this case, one has additional states in the
form of transverse twisted sectors.  This will be shown to create
a problem with state interpretation.

In addition to the transverse untwisted states defined by
(\ref{eqn:TransverseAnnulusWT}), one now has
\begin{eqnarray}
\tilde{{\cal
A}}&=&\frac{2^{-5}}{16}\bigg{\{}\bigg{(}N_o^2v_1v_2v_3(W^1W^2W^3+
W^1_{n+\frac{1}{2}}W^2_{n+\frac{1}{2}}W^3_{n+\frac{1}{2}})\nn\\
&&+\frac{v_k}{2v_lv_s}D^2_{ko}W^kP^lP^m\frac{\big{(}1+(-1)^{m_l+m_s}\big{)}}{2}
\bigg{)}T_{oo}\nn\\
&&+\bigg[M_1N_k^2v_k(W^k+W^k_{n+\frac{1}{2}})\nn\\
&&+M_2D_{kk}^2v_kW^k+M_3D^2_{lk}\frac{P^k}{v_k}\bigg]\tilde{T}_{ko}
{\bigg{(}\frac{\eta}{\theta_4}\bigg{)}}^2\nn\\
&&+2N_oD_{ko}v_kW^k\tilde{T}_{ok}^{(\epsilon_k)}{\bigg{(}\frac{2\eta}
{\theta_2}\bigg{)}}^2\nn\\
&&+M_4N_kD_{kk}v_kW^k\tilde{T}_{kk}^{(\epsilon_k)}{\bigg{(}\frac{\eta}
{\theta_3}\bigg{)}}^2\nn\\
&&+M_5N_lD_{kl}\tilde{T}_{lk}^{(\epsilon_k)}\frac{8{\eta}^3}{\theta_2
\theta_3\theta_4}\nn\\
&&+D_{ko}D_{lo}\frac{P^s}{v_s}\frac{\big{(}1+(-1)^{m_s}\big{)}}{2}
\tilde{T}_{os}^{(\epsilon_k\epsilon_l)}{\bigg{(}\frac{2\eta}{\theta_2}
\bigg{)}}^2\nn\\
&&+M_{6}D_{km}D_{lm}\frac{P^m}{v_m}\tilde{T}_{mm}^{(\epsilon_k
\epsilon_l)}{\bigg{(}\frac{\eta}{\theta_3}\bigg{)}}^2\nn\\
&&+M_{7}D_{kk}D_{lk}\tilde{T}_{km}^{(\epsilon_k\epsilon_l)}
\frac{8{\eta}^3}{\theta_2\theta_3\theta_4}\bigg{\}}.
\end{eqnarray}
The coefficients $M_i$ are determined from the origin of the
twisted sector.  The origin of such sectors must reflect the fixed
point multiplicity of its constituent brane couplings.

The $N_g$ term fills all compact and non-compact dimensions and
thus has the coefficient $v_k$.  With the volume $v_k$ being
provided by the remaining compact direction that is not acted on
by an orbifold.  When considering the factors involved with terms
like $D_{kl}$, one proceeds understanding the terminology that $l$
represents the fixed point configuration of $T^2_{45} \times
T^2_{67} \times T^2_{89}$ and $k$ represents whether the brane is
wrapped or transverse. For example, $D_{gf}$ has fixed points in
the first and third torus corresponding to $f$. The index $g$
implies that $D_{gf}$ brane is wrapped around the first tori and
is transverse to the second and third, consistent with the
representation $g=(+,-,-)$. Hence, it \textit{sees} four fixed
points.

Looking at the $g$-twisted sector, this has a total of sixteen
fixed points, four located in each of the second and third tori.
Under the operation of the shift, half are identified. The
independent fixed points are as in table
\ref{tab:shiftedFixedPoints}.

For the $g$-twisted sector, one has terms in the annulus as
\begin{eqnarray}
\tilde{{\cal
A}}^g=&\frac{2^{-5}}{16}&\bigg{\{}\bigg[(M_1N_g^2+M_2D_{gg}^2)v_1+
M_3D^2_{lg}\frac{1}{v_1}\bigg]\tilde{T}_{go}\nn\\ \nn\\
&&+M_4N_gD_{gg}v_1\tilde{T}_{gg}^{(\epsilon_1)}\nn\\ \nn\\
&&+4M_5N_gD_{kg}\tilde{T}_{gk}^{(\epsilon_k)}\nn\\ \nn\\
&&+M_{6}D_{kg}D_{lg}\frac{1}{v_1}\tilde{T}_{gg}^{(\epsilon_k\epsilon_l)}
\nn\\ \nn\\
&&+4M_{7}D_{gg}D_{lg}\tilde{T}_{gm}^{(\epsilon_k\epsilon_l)}\bigg\}.\nn\\
\end{eqnarray}

All brane types $N_g$, $D_{gg}$, $D_{fg}$ and $D_{hg}$ see the
fixed point $(0,0;0,0)$, and therefore arrange into a perfect
square which multiplicity 1.  The arguments set out in the simpler
${\cal N}=2$ model with regard to the wrapping of $D5$ branes is
generalized here with the inclusion of three distinct types.  The
counting of their fixed point occupation is then a little more
complicated.  $N_g$ and $D_{hg}$ see the fixed points
$(0,0,\frac{1}{2})$, $(0,0;0,\frac{1}{2}$ and
$(0,0;\frac{1}{2},\frac{1}{2})$, which correspond to their own
perfect square with multiplicity 3.  The coefficients of the $N_g$
and $D_{hg}$ terms are 1 and 2 respectively, which can easily be
seen by reference to (\ref{eqn:FixedPointSummary}).  Similar holds
for the square of $N_g$ and $D_{fg}$. The remaining fixed points
are taken into account by $N_g$ alone.

The resulting perfect square structure for the $\tau_{gl}$ with
orbifold element $g$ is
\begin{eqnarray}
\tilde{{\cal
A}}^g_{o}&=&2\times\frac{2^{-5}}{16}\bigg{\{}{\big{(}\sqrt{v_1}N_g+
4s_1\sqrt{v_1}D_{gg}+2s_2\frac{1}{\sqrt{v_1}}D_{fg}+2s_3
\frac{1}{\sqrt{v_1}}D_{hg}\big{)}}^2\nn\\ \nn\\
&&+3{\big{(}\sqrt{v_1}N_g+2s_4\frac{1}{\sqrt{v_1}}D_{fg}\big{)}}^2
+3{\big{(}\sqrt{v_1}N_g+2s_5\frac{1}{\sqrt{v_1}}D_{hg}\big{)}}^2+
v_1N_g^2\bigg{\}},\nn\\
\end{eqnarray}
where the signs $s_i$ are completely determined by the orbifold
direction $o,g,f$ and $h$ within the $g$ twist and the sector that
is considered, $NS$ or $R$.  The overall factor of 2 is to account
for the multiplicity of the shifted fixed points in the same
fashion as for the $\bb{Z}_2\times \bb{Z}_2$ (shift) case.

With the aid of the identity
\begin{eqnarray}
\theta_2 \theta_3 \theta_4 = 2\eta^3,\nn
\end{eqnarray}
the transverse annulus is now seen to be
\begin{eqnarray}
\tilde{{\cal
A}}=&\frac{2^{-5}}{16}&\bigg{\{}\bigg{(}N_o^2v_1v_2v_3
(W^1W^2W^3+W^1_{n+\frac{1}{2}}W^2_{n+\frac{1}{2}}
W^3_{n+\frac{1}{2}})\nn\\
&&+\frac{v_k}{2v_lv_s}D^2_{ko}W^kP^lP^s
\frac{\big{(}1+(-1)^{m_l+m_s}\big{)}}{2}\bigg{)}T_{oo}\nn\\
&&+2\times2\bigg{[}N_k^2v_k(W^k+W^k_{n+\frac{1}{2}})\nn\\
&&+2D_{kk}^2v_kW^k+2D^2_{lk}\frac{P^k}{v_k}\bigg{]}
\tilde{T}_{ko}{\bigg{(}\frac{2\eta}{\theta_4}\bigg{)}}^2\nn\\
&&+2N_oD_{ko}v_kW^k\tilde{T}_{ok}^{(\epsilon_k)}
{\bigg{(}\frac{2\eta}{\theta_2}\bigg{)}}^2\nn\\
&&+2\times2N_kD_{kk}v_kW_k\tilde{T}_{kk}^{(\epsilon_k)}
{\bigg{(}\frac{2\eta}{\theta_3}\bigg{)}}^2\nn\\
&&+2\times4N_lD_{kl}\tilde{T}_{lk}^{(\epsilon_k)}
\frac{8{\eta}^3}{\theta_2\theta_3\theta_4}\nn\\
&&+D_{ko}D_{lo}\frac{P^s}{v_s}\frac{\big{(}1+(-1)^{m_s}\big{)}}{2}
\tilde{T}_{os}^{(\epsilon_k\epsilon_l)}{\bigg{(}\frac{2\eta}{\theta_2}
\bigg{)}}^2\nn\\
&&+2D_{km}D_{lm}\frac{P^m}{v_m}\tilde{T}_{mm}^{(\epsilon_k\epsilon_l)}
{\bigg{(}\frac{2\eta}{\theta_3}\bigg{)}}^2\nn\\
&&+2\times4D_{kk}D_{lk}\tilde{T}_{km}^{(\epsilon_k\epsilon_l)}
\frac{8{\eta}^3}{\theta_2\theta_3\theta_4}\bigg{\}}.\nn\\
\end{eqnarray}
With corresponding direct channel
\begin{eqnarray}\label{eqn:dirannorigin}
{\cal
A}=&\frac{1}{16}&\bigg{\{}\bigg{(}N_o^2P^1P^2P^3\big(1+(-1)^{m_1+m_2+m_3}\big)
\nn\\
&&+\frac{1}{2}\frac{D^2_{ko}}{2}P^k(W^lW^m+W^l_{n+\frac{1}{2}}
W^m_{n+\frac{1}{2}}\bigg{)}T_{oo}\nn\\
&&+\bigg{[}N_k^2P^k\big(1+(-1)^{m_k}\big)\nn\\
&&+2D_{kk}^2P^k+2D^2_{lk}W^k\bigg{]}T_{ok}{\bigg{(}\frac{2\eta}{\theta_2}
\bigg{)}}^2\nn\\
&&+2N_oD_{ko}P^k
T_{ko}^{(\epsilon_k)}{\bigg{(}\frac{\eta}{\theta_4}\bigg{)}}^2\nn\\
&&-2\times2N_kD_{kk}P^kT_{kk}^{(\epsilon_k)}{\bigg{(}\frac{\eta}{\theta_3}
\bigg{)}}^2\nn\\
&&+2\times2i(-1)^{k+l}N_lD_{kl}T_{kl}^{(\epsilon_k)}
\frac{2{\eta}^3}{\theta_2\theta_3\theta_4}\nn\\
&&+\frac{1}{2}D_{ko}D_{lo}(W^m+W^m_{n+\frac{1}{2}})
T_{mo}^{(\epsilon_k\epsilon_l)}{\bigg{(}\frac{\eta}{\theta_4}\bigg{)}}^2
\nn\\
&&-2\times D_{km}D_{lm}W^mT_{mm}^{(\epsilon_k\epsilon_l)}{\bigg{(}
\frac{\eta}{\theta_3}\bigg{)}}^2\nn\\
&&+2\times2i(-1)^{m+k}D_{kk}D_{lk}T_{mk}^{(\epsilon_k\epsilon_l)}
\frac{2{\eta}^3}{\theta_2\theta_3\theta_4}\bigg{\}}.\nn\\
\end{eqnarray}

It is here that consistent particle interpretation does not occur.
The inconsistency is generated in the $D5_i-D5_j$ (for $i\neq j$)
sector.  All other sectors give rise to the proper massless and
massive counting.  Moreover, the problem exists in the twisted
sector that has to symmetrize by itself, as the Mobius has only
untwisted sectors present.  The Mobius has the same form as for
the case without discrete torsion with the appropriate change of
the charges $\epsilon_k$ so as to allow $\epsilon=-1$.

We look at the particular case of $\epsilon=(+,+,-)$.  To begin
with, one must define the Chan-Paton charge parameterization,
which is given in table \ref{tab:ModelCharges}. The relative signs
and factors of $i$ are fixed by requiring that the spectrum is
real and that the vector multiplet be in the oriented $n\bar{n}$
representation, and thus absent from the Mobius.  In addition the
scaling factors of two for $D_{ko}$ charges are necessary, and
induced by the shift, to give proper integer particle
interpretation.  All sectors that involve a coupling to a $D9$
brane are consistent.

\begin{table}[!ht]
\begin{center}
\begin{tabular}{|llllll|}\hline
$N_o$ & $=$ & $(n+m+\bar{n}+\bar{m})$, & $N_g$ & $=$ &
$i(n+m-\bar{n}-\bar{m})$
\\
$N_f$ & $=$ & $i(n-m-\bar{n}+\bar{m})$, & $N_h$ & $=$ &
$(n-m+\bar{n}-\bar{m})$\\
$D_{go}$ & $=$ & $2(o_1+g_1+\bar{o}_1+\bar{g}_1)$, & $D_{fo}$ &
$=$ &
$2(o_2+g_2+\bar{o}_2+\bar{g}_2)$\\
$D_{ho}$ & $=$ & $2(a+b+c+d)$, & $D_{gg}$ & $=$ &
$i(o_1+g_1-\bar{o}_1-\bar{g}_1)$\\
$D_{ff}$ & $=$ & $i(o_2+g_2-\bar{o}_2-\bar{g}_2)$, & $D_{hh}$ &
$=$ &
$a-b-c+d$\\
$D_{gf}$ & $=$ & $o_1-g_1+\bar{o}_1-\bar{g}_1$, & $D_{gh}$ & $=$ &
$-i(o_1-g_1-\bar{o}_1+\bar{g}_1)$\\
$D_{fg}$ & $=$ & $o_2-g_2+\bar{o}_2-\bar{g}_2$, & $D_{fh}$ & $=$ &
$i(o_2-g_2-\bar{o}_2+\bar{g}_2)$\\
$D_{hg}$ & $=$ & $a+b-c-d$, & $D_{hf}$ & $=$ & $a-b+c-d$\\\hline
\end{tabular}
\end{center}
\caption{$\epsilon=(1,1,-1)$ Model
Charges}\label{tab:ModelCharges}
\end{table}
With this parameterization, the untwisted sector provides
consistent massless spectrum as
\begin{eqnarray}\label{egn:AnplusMobzero}
{\cal A}_o+{\cal
M}_o&=&(n\bar{n}+m\bar{m}+g_1\bar{g}_1+o_1\bar{o}_1+o_2\bar{o}_2+g_2\bar{g}_2)
\tau_{oo}\nn\\ \nn\\
&&+(n\bar{m}+m\bar{n}+o_1\bar{g}_1+g_1\bar{o}_1+ab+cd)\tau_{og}\nn\\ \nn\\
&&+(nm+\bar{n}\bar{m}+o_2\bar{g}_2+g_2\bar{o}_2+ac+bd)\tau_{of}\nn\\ \nn\\
&&+(\bar{o}_1\bar{g}_1+o_1g_1+o_2g_2+\bar{o}_2\bar{g}_1+ad+bc)\tau_{oh}\nn\\
\nn\\
&&+\frac{\big(a(a+1)+b(b+1)+c(c+1)+d(d+1)\big)}{2}\tau_{oo}^{NS}\nn\\ \nn\\
&&+\frac{\big(a(a-1)+b(b-1)+c(c-1)+d(d-1)\big)}{2}\tau_{oo}^{R}\nn\\ \nn\\
&&+\frac{\big(o_2(o_2-1)+g_2(g_2-1)+\bar{o}_2(\bar{o}_2-1)+\bar{g}_2
(\bar{g}_2-1)\big)}{2}\tau_{og}\nn\\ \nn\\
&&+\frac{\big(o_1(o_1-1)+g_1(g_1-1)+\bar{o}_1(\bar{o}_1-1)+\bar{g}_1
(\bar{g}_1-1)\big)}{2}\tau_{of}\nn\\ \nn\\
&&+\frac{\big(n(n-1)+m(m-1)+\bar{n}(\bar{n}-1)+
\bar{m}(\bar{m}-1)\big)}{2}\tau_{oh}.\nn\\
\end{eqnarray}
Indeed, the twisted massless spectrum also gives rise to a
consistent particle interpretation in all sectors $g$, $f$ and
$h$.  However, for the $n+\frac{1}{2}$ massive $D5_i-D5_j$ sector,
one has
\begin{eqnarray}
\frac{1}{2}D_{fo}D_{lo}W^m_{n+\frac{1}{2}}
T_{mo}^{(\epsilon_k\epsilon_l)}{\bigg{(}\frac{\eta}{\theta_4}\bigg{)}}^2.
\nn\\
\end{eqnarray}
Any combination of the generators $g$, $f$ and $h$ will map
$W_{n+\frac{1}{2}}$ to $W_{\pm(n+\frac{1}{2})}$, this winding
tower will then have a degeneracy of two.  Taking into account the
interchange counting $k\leftrightarrow l$ and the rescaling
defined in table \ref{tab:ModelCharges}, the end result is a state
with numerical coefficient of $\frac{1}{2}$.

The same term occurred in the model without discrete torsion in
eqn. (\ref{eqn:AnnulusRescaled}).  In that case, it did not cause
any inconsistency because of the generic rescaling $N\rightarrow
2N$ and $D\rightarrow 2D$ in addition to the rescaling induced by
the freely acting shift.  In the models with discrete torsion,
such a rescaling is taken into account (for the integer massive
and massless levels) by the presence of the breaking terms $N_k,
D_{kk},\ldots$.  For example, the character $\tau_{oo}$ coupling
to $D5$ branes includes the terms
$2D_{go}^2+2D_{gg}^2+2D_{gf}^2+2D_{gh}^2=
16(o_1\bar{o}_1+\ldots)$.  So, to introduce additional rescaling
of the $D5$ branes would produce an over counting at the massless
level.

In the transverse annulus, one has the freedom to introduce Wilson
lines via phases of the form $e^{2\pi i \alpha}$, for $\alpha\in
(0,1)$. If such phases are introduced, the resulting amplitude
must respect symmetrization in the direct channel and the
corresponding terms in the transverse annulus must exist in the
torus. Introducing phases in the group of $D5-D5$ terms that must
symmetrize together as
\begin{eqnarray}
&&\frac{1}{2}D_{ko}D_{lo}(W^m+W^m_{n+\frac{1}{2}})
T_{mo}^{(\epsilon_k\epsilon_l)}{\bigg{(}\frac{\eta}{\theta_4}\bigg{)}}^2
\nn\\
&&-2\times D_{km}D_{lm}W^mT_{mm}^{(\epsilon_k\epsilon_l)}{\bigg{(}
\frac{\eta}{\theta_3}\bigg{)}}^2\nn\\
&&+2\times2i(-1)^{m+k}D_{kk}D_{lk}T_{mk}^{(\epsilon_k\epsilon_l)}
\frac{2{\eta}^3}{\theta_2\theta_3\theta_4}.\nn\\
\end{eqnarray}
only leads to amplitudes that violate these requirements.  The
first choice for a phase in the transverse channel amplitude would
be of the form
\begin{eqnarray}
D_{fo}D_{lo}P^s\frac{\big(1+(-1)^{m_s^1+\beta m_s^2}\big)}{2}
\quad\overrightarrow{S}\quad
\frac{1}{2}D_{fo}D_{lo}\big(W^s+W_{n_1+\frac{1}{2},n_2+\frac{\beta}{2}}^s\big).
\nn\\
\end{eqnarray}
Where $m_s^1$ and $m_s^2$ are the momentum quantum numbers on the
first and second directions of a given compact direction. For
$\beta=1$, the momentum lattice in the transverse channel has
expanded form
\begin{eqnarray}
P^s_{2m_1,2m_2}+P^s_{2m_1+1,2m_2+1}
\nn\\
\end{eqnarray}
which includes odd states that do not exist in the torus.  For
$\beta\neq 1$, the projection which leads to the counting of even
momentum states, as required by the torus, no longer exists unless
$\beta=0~{\rm mod}~2$.  Therefore, the introduction of a phase for
this term has to be one which multiplies the whole lattice
expression.  Doing this will however lift the massless spectrum in
the direct channel amplitude.


\section{Conclusions}


The spectra of freely acting orbifolds with non-freely acting
winding and or Kaluza Klein shifts have been exhaustively studied
in \cite{AADS} as (shift) orbifolds. The inclusion of shift
operators within the generators of the $\bb{Z}_2\times \bb{Z}_2$
generators leads to (in the cases with two $D5$ branes, and some
models with only one $D5$ brane) richer geometries.  In
particular, such cases involve shifted fixed points, which give
rise to unique massive lattices of the form $W_{n+\frac{1}{4}}$ in
the direct channel annulus. In addition, the arrangement of shifts
within the $\bb{Z}_2\times \bb{Z}_2$ generators imposes
restrictions on the number of distinct $D5$ branes that can exist.
The models defined by $\bb{Z}_2\times \bb{Z}_2$ (shift)
projections that essentially exhaust all interesting
configurations are defined \cite{AADS} by
\begin{eqnarray}\label{eqn:(shift)orbifolds}
\sigma_1(\delta_1,\delta_2,\delta_3)=\left(\matrix{ \delta_1 &
-\delta_2 & -1 \cr -1 & \delta_2 & -\delta_3 \cr -\delta_1 & -1 &
\delta_3 \cr }\right),\quad
\sigma_2(\delta_1,\delta_2,\delta_3)=\left(\matrix{ \delta_1 & -1
& -1 \cr -1 & \delta_2 & -\delta_3 \cr -\delta_1 & -\delta_2 &
\delta_3 \cr }\right).
\end{eqnarray}
The parameters $\delta_i$ are winding or momentum shifts. Wherever
a $\delta$ operation exists in a column, the corresponding brane
is eliminated.

In the freely acting shift models, all $D5$ branes are allowed to
exist but have a more conventional geometry with regard to their
relative placements.

The reduction in overall closed spectral content in the (shift)
orbifold cases arises from the action of the shift on the fixed
point counting. This is especially evident in that such (shift)
orbifolds do not allow contributions from $\bb{Z}_2\times
\bb{Z}_2$ orbits that lie outside $S$ and $T$ transformations on
the principle orbits $(o,o),(o,g),(o,f)$ and $(o,h)$ (as
illustrated in \ref{app:Boundaries}).

The freely acting Kaluza Klein shift orbifold construction clearly
allows such orbits since there can exist pure orbifold twisted
states. Furthermore, this arrangement increases the massive
spectral content of the twisted sector of the parent
$\bb{Z}_2\times \bb{Z}_2$ orbifold modulated torus.  In addition,
the conventional massive and massless lattice states in the
twisted sector of the parent torus are maintained.

With the inclusion of independent orbits, one has a class of
models which exhibit possible scenarios of supersymmetric or
non-supersymmetric models (according to the sign freedom
associated with the independent modular orbits). The cases without
discrete torsion which include $(+,+,+)$, $(+,-,-)$, $(-,+,-)$ and
$(-,-,+)$ lead to fully consistent amplitudes with ${\cal N}=1$
supersymmetry in the open sector for the $(+,+,+)$ model and
broken supersymmetry in the others.  This brane supersymmetry
breaking is associated with those branes aligned with the
directions corresponding to $\epsilon_k=-1$.

There is however an unresolved problem of consistent particle
interpretation for cases with discrete torsion for $n+\frac{1}{2}$
massive modes stretched between $D5$ branes of any type.

The sign $\epsilon$ associated with the inclusion of the
additional independent orbits is a freedom of choice.  There is no
mechanism outlined yet that guides the choice of which model is
preferred. However, for the subclasses of say $\epsilon=+1$, three
of the models are related, these are given by $(+,-,-)$, $(-,+,-)$
and $(-,-,+)$. Similarly occurs for the subclasses of
$\epsilon=-1$, so one has an overall set of four independent
models which display different phenomenology. In contrast to the
(shift) orientifold models, although the closed spectrum is not as
rich in such cases, they do eliminate this freedom.

In many of the possible arrangements of shifts, the corresponding
torus amplitude allows the propagation of twisted states in the
transverse annulus.  The cancellation of the twisted tadpoles then
allow the existence of such sectors with breaking terms that have
Chan-Paton parameterizations that lead to unitary groups.
Moreover, models defined by (\ref{eqn:(shift)orbifolds}), also
display open spectra with mixed orthogonal and unitary gauge group
types for both the $D9$ and $D5$ sectors.  In the case of the
freely acting shift, the form of the gauge group is confined to
the representation provided by the parent $\bb{Z}_2\times
\bb{Z}_2$ group.


\section{Acknowledgments}


We would like to thank Carlo Angelantonj,
Emilian Dudas and Jihad Mourad for useful
discussions. AF would like to thank the Orsay theory group and
LPTENS for hospitality in the initial phase of this work. This
work is supported in part by the Royal Society and PPARC.

\appendix

\section{Shift Action on Mass}\label{app:massshift}
Here, we show the explicit action of the shift on mass after an
$S$ transformation.  The compact form (\ref{eqn:compact}) is
written in contour form
\begin{eqnarray}
(-1)^{m_k}\Lambda_{m_k,n_k}=\frac{1}{2\pi
i}\oint_C\frac{d^2z}{e^{2\pi i z}-1}e^{i\pi \big
(z+\frac{2i\tau_2z^2}{R^2}+2\pi z n \tau_1+i\frac{2\tau_2 n^2
R^2}{4}\big)}.\nn
\end{eqnarray}
For the contours, take
\begin{eqnarray}
(Imz>0),&&C_1~\leftrightarrow~\frac{1}{e^{2\pi i
z}-1}=-\sum^{\infty}_{\tilde{m}}e^{2\pi i\tilde{m}z}\nn\\
(Imz<0),&&C_2~\leftrightarrow~e^{2\pi i z}\frac{1}{e^{2\pi i
z}-1}=\sum^{-1}_{\tilde{m}=-\infty}e^{2\pi i\tilde{m}z}.\nn
\end{eqnarray}
The two contour integrals become
\begin{eqnarray}
\frac{1}{2\pi
i}\int_{-\infty}^{\infty}dz\sum_{\tilde{m}=-\infty}^{\infty}
e^{\big\{-\frac{2\pi
\tau_2}{R^2}\big(z-\frac{iR^2}{2\tau_2}(\tilde{m}+
\frac{1}{2}+n\tau_1)\big)^2+\frac{\pi
R^2}{2\tau_2}(\tilde{m}+\frac{1}{2}+n\tau_1)^2-\frac{\pi\tau_2
n^2R^2}{2}\big\}}.\nn
\end{eqnarray}
By virtue of the gaussian integral, this is thus represented as
\begin{eqnarray}
\frac{R}{\sqrt{2\tau_2}}\sum_{\tilde{m}=-\infty}^{\infty}e^{-\frac{\pi
R^2}{2\tau_2}|\tilde{m}+\frac{1}{2}+n\tau|^2}~\rightarrow~
\frac{R}{\sqrt{2\tau_2}}\sum_{\tilde{m}=-\infty}^{\infty}e^{-\frac{\pi
R^2}{2\tau_2}|n+(\tilde{m}+\frac{1}{2})\tau|^2}\nn
\end{eqnarray}
after $S$ transformation which acts on the measure components as
\begin{eqnarray}
S:\left(\matrix{ \tau_1 \cr \tau_2 }\right) ~\rightarrow~
\left(\matrix{-\frac{\tau_1}{|\tau|^2} \cr
\frac{\tau_2}{|\tau|^2}\nn }\right).
\end{eqnarray}
This then shows that the roles $\tilde{m}$ and $n$ are
interchanged as winding and Kaluza-Klein respectively. This then
shows the resulting shift in winding induced by the S transform
involving a Kaluza-Klein phase.

\section{General Mobius Origin for the $A_1$ Shift}\label{app:moborigin}

This is the Mobius origin for all classes of models.  It is
provided as a reference to show how the choice of different
classes results in the change of gauge structure through the signs
that are provided by a given model class.
\begin{eqnarray}
{\cal
M}_o=&-\frac{1}{8}&\bigg{\{}\bigg(2N_o(1-\epsilon_1-\epsilon_2-\epsilon_3)-
D_{go}(1-\epsilon_1+\epsilon_2+\epsilon_3)\nn\\
&&-D_{fo}(1+\epsilon_1-\epsilon_2+\epsilon_3)-D_{ho}(1+\epsilon_1+
\epsilon_2-\epsilon_3)\bigg)\hat{\tau}_{oo}^{NS}\nn\\
&&-\bigg(2N_o(1-\epsilon_1-\epsilon_2-\epsilon_3)-D_{go}\epsilon_1(1-
\epsilon_1+\epsilon_2+\epsilon_3)\nn\\
&&-D_{fo}\epsilon_2(1+\epsilon_1-\epsilon_2+\epsilon_3)-
D_{ho}\epsilon_3(1+\epsilon_1+\epsilon_2-\epsilon_3)\bigg)
\hat{\tau}_{oo}^{R}\bigg{\}}\nn\\
&&+\bigg(2N_o(1-\epsilon_1+\epsilon_2+\epsilon_3)-
D_{go}(1-\epsilon_1-\epsilon_2-\epsilon_3)\nn\\
&&-D_{fo}(-1-\epsilon_1-\epsilon_2+\epsilon_3)-
D_{ho}(-1-\epsilon_1+\epsilon_2-\epsilon_3)\bigg)\hat{\tau}_{og}^{NS}\nn\\
&&-\bigg(2N_o(1-\epsilon_1+\epsilon_2+\epsilon_3)-
D_{go}\epsilon_1(1-\epsilon_1-\epsilon_2-\epsilon_3)\nn\\
&&-D_{fo}\epsilon_2(-1-\epsilon_1-\epsilon_2+\epsilon_3)-
D_{ho}\epsilon_3(-1-\epsilon_1+\epsilon_2-\epsilon_3)\bigg)
\hat{\tau}_{og}^{R}\bigg{\}}\nn\\
&&+\bigg(2N_o(1+\epsilon_1-\epsilon_2+\epsilon_3)-
D_{go}(-1-\epsilon_1-\epsilon_2+\epsilon_3)\nn\\
&&-D_{fo}(1-\epsilon_1-\epsilon_2-\epsilon_3)-
D_{ho}(-1+\epsilon_1-\epsilon_2-\epsilon_3)\bigg)
\hat{\tau}_{of}^{NS}\nn\\
&&-\bigg(2N_o(1+\epsilon_1-\epsilon_2+\epsilon_3)-
D_{go}\epsilon_1(-1-\epsilon_1-\epsilon_2+\epsilon_3)\nn\\
&&-D_{fo}\epsilon_2(1-\epsilon_1-\epsilon_2-\epsilon_3)-
D_{ho}\epsilon_3(-1+\epsilon_1-\epsilon_2-\epsilon_3)\bigg)
\hat{\tau}_{of}^{R}\bigg{\}}\nn\\
&&+\bigg(2N_o(1+\epsilon_1+\epsilon_2-\epsilon_3)-
D_{go}(-1-\epsilon_1+\epsilon_2-\epsilon_3)\nn\\
&&-D_{fo}(-1+\epsilon_1-\epsilon_2-\epsilon_3)-
D_{ho}(1-\epsilon_1-\epsilon_2-\epsilon_3)\bigg)
\hat{\tau}_{oh}^{NS}\nn\\
&&-\bigg(2N_o(1+\epsilon_1+\epsilon_2-\epsilon_3)-
D_{go}\epsilon_1(-1-\epsilon_1+\epsilon_2-\epsilon_3)\nn\\
&&-D_{fo}\epsilon_2(-1+\epsilon_1-\epsilon_2-\epsilon_3)-
D_{ho}\epsilon_3(1-\epsilon_1-\epsilon_2-\epsilon_3)\bigg)
\hat{\tau}_{oh}^{R}\nn
\end{eqnarray}

\section{$\bb{Z}_2\times \bb{Z}_2$ Boundary Operators}\label{app:Boundaries}

The $\bb{Z}_2\times \bb{Z}_2$ generators including the identity
lead to $16=4\times 4$ distinct boundary conditions on the two
dimensional sheet.  These are portrayed in fig. \ref{app:Blocks}.
The shaded blocks represent those which are not connected to the
unshaded ones by modular invariance, or $S$ and $T$ transforms.

\begin{figure}[!h]
\centerline{\epsfxsize 4.0 truein \epsfbox {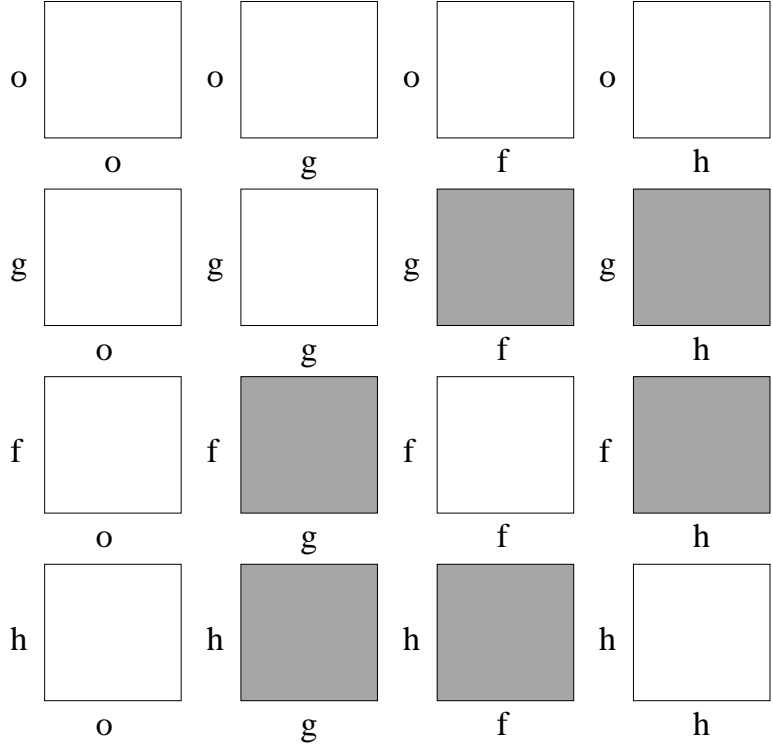}}
\caption{Distinct boundary sets in the $\bb{Z}_2\times \bb{Z}_2$}
\label{app:Blocks}
\end{figure}


\vfill\eject

\bigskip
\medskip

\bibliographystyle{unsrt}

\end{document}